\documentclass[12pt]{article}
\usepackage[
top = 2.5cm, 
bottom = 2.5cm,  
left = 2.55cm,  
right = 2.55cm]{geometry}

\usepackage{bbm}
\usepackage{overpic}
\usepackage{subfigure}
\usepackage{caption}
\usepackage{mathtools}
\usepackage{latexsym} 
\usepackage{verbatim}  
\usepackage{tikz}
\usetikzlibrary{matrix}
\usepackage{color}
\usepackage{graphicx,amssymb,amsfonts,amsmath,amssymb,amscd,amstext, mathrsfs}
\usepackage{graphicx}
\usepackage{dsfont}
\usepackage{xcolor}
\usepackage{amsmath}
\usepackage{empheq}
\usepackage{cite}
\usepackage{mmacells}

\numberwithin{equation}{section}

\newlength\dlf

\def\l{\lambda}

\def\CV{\mathcal{V}}

\newcommand{\bw}{\begin{widetext}}
\newcommand{\ew}{\end{widetext}}
\newcommand{\bea}{\begin{eqnarray}}
\newcommand{\eea}{\end{eqnarray}}
\newcommand{\be}{\begin{equation}}
\newcommand{\ee}{\end{equation}}
\newcommand{\nn}{\nonumber}

\renewcommand{\bar}[1]{\overline{#1}}

\renewcommand{\hat}[1]{\widehat{#1}}

\newcommand{\<}{\langle}
\renewcommand{\>}{\rangle}

\newcommand{\CO}{\mathcal{O}}

\newcommand{\kvec}{ {\boldsymbol{k}} }

\newcommand{\Dmax}{\Delta_{\max}}

\DeclareFontShape{OT1}{cmr}{mx}{n}{<->cmr10}{}
\newcommand{\titlefont}{\fontseries{mx}\selectfont}

\usepackage{jheppub}
\begin{document}

\begin{titlepage}

\begin{flushright} 
\end{flushright}

\begin{center} 

\vspace{0.35cm}

{\fontsize{20.5pt}{25pt}
{\titlefont 
Toolkit for General 2d Scalar  Potential in LCT
}}

\vspace{1.6cm}  

{{A. Liam Fitzpatrick$^1$,  Emanuel Katz$^1$,  Yuan Xin$^{2}$}}

\vspace{1cm} 

{{\it
$^1$Department of Physics, Boston University, 
Boston, MA  02215, USA
\\
\vspace{0.1cm}
$^2$Department of Physics, CMU, Pittsburgh, PA 15213, USA \\
\vspace{0.1cm}
}}
\end{center}
\vspace{1.5cm}

{\noindent 
We present efficient algorithms for obtaining the Hamiltonian in Lightcone Conformal Truncation (LCT) for a 2d scalar field with a generic potential.  We apply this method to the sine-Gordon and sinh-Gordon models in $1+1d$, and find precise agreement with integrability results when the scaling dimension $\Delta$ of the deforming cosine/cosinh potential is in the range $ \Delta \le 1$.  The agreement provides additional evidence for a recent conjecture for how to compute the effective lightcone Hamiltonian in this class of models.
In addition, to high precision, we provide the first direct confirmation for the conjectured self-duality of the sinh-Gordon model ($\Delta<0)$, which relates $\Delta \leftrightarrow 4/\Delta$.   As the dimension approaches  the upper limit $\Delta=1$ from below, we show analytically that the Hamiltonian matrix elements exactly reproduce those of a free Majorana fermion, demonstrating how bosonization is manifested in the LCT basis.  We comment on the possible extension of the approach to $\Delta >1$.
}

\end{titlepage}

\tableofcontents

%%%%%%%%%%%%%%%%%%%%%%
%%%%%%%%%%%%%%%%%%%%%%
\newpage

\newpage
\section{Introduction and Summary} 
 
Hamiltonian truncation is a method for doing strongly coupled calculations in Quantum Field Theory (QFT), or any infinite dimensional quantum system, by numerically diagonalizing the Hamiltonian of the theory on a finite-dimensional subspace \cite{Hornbostel:1988fb,Yurov:1991my,Hogervorst:2014rta,Rychkov:2014eea,Coser:2014lla,Rychkov:2015vap,Delouche:2023wsl,Elias-Miro:2017xxf,Hogervorst:2018otc,Elias-Miro:2020qwz,Cohen:2021erm,Tilloy:2021yre,EliasMiro:2022pua,Dempsey:2022uie,Horvath:2022zwx,Lencses:2022ira,Schmoll:2023eez,Lencses:2024wib,Katsevich:2024sov,Delouche:2024tjf}. Lightcone Conformal Truncation (LCT)\cite{Anand:2020qnp,Anand:2020gnn,Delacretaz:2018xbn} is a particular implementation of this idea, where one works in lightcone quantization and uses a truncated subspace motivated by the conformal symmetry of the ultraviolet (UV) of the theory.   The numeric efficiency of the method is greatest in lower dimensions, and at this point large classes of  2d QFTs can be studied at high precision using currently available techniques \cite{Katz:2013qua,Katz:2014uoa,Anand:2017yij,Fitzpatrick:2018xlz,Fitzpatrick:2019cif,Anand:2021qnd,Chen:2021bmm,Chen:2021pgx,Chen:2023glf,Fitzpatrick:2023aqm}. 

 However, despite the relative versatility of this approach for 2d QFTs, it has mostly been applied to a fairly small number of models.  Part of the reason for this is that each new theory application takes time to develop algorithms to compute the corresponding Hamiltonian efficiently in practice, even in the case where the UV CFT is a free theory and can be exactly solved in principle.  On the other hand, when we think of a Lagrangian theory with only relevant interactions, there are only a handful of different possible types of interactions available, such as Yukawa interactions, gauge interactions, and potentials made from products of scalar fields.  It seems both quite feasible and useful to implement the existing methods on this entire class of theories, which could be a valuable tool for quickly investigating questions of strongly coupled physics in a 2d setting.
 
In this paper, we introduce an implementation that covers a large subset of such theories, namely a single real scalar field with an arbitrary $\mathbb{Z}_2$-even $V(\phi)$ potential.  Unlike in $d>2$, any $\phi^n$ interaction is relevant, providing a huge space of interesting RG flows into the infrared (IR); these include  RG flows to all the unitary $c<1$ minimal models, for instance. Code implementing this approach is available online at \href{https://github.com/andrewliamfitz/LCT}{\texttt{https://github.com/andrewliamfitz/LCT}}.

\subsection{Summary of Main Results}

 To demonstrate the method, we apply it to the sine-Gordon and sinh-Gordon models.   
Both the sine-Gordon and sinh-Gordon model can be thought of as the deformation
\begin{equation}
\delta {\mathcal{L} } = -\frac{\lambda_{\rm LC}}{2\pi} \cos(\beta \phi),
\end{equation}
where $\lambda_{\rm LC}$ is the lightcone quantization `coupling', and $\beta$ is real (imaginary) for sine-Gordon (sinh-Gordon).  The effective lightcone quantization coupling $\lambda_{\rm LC}$ is related to the equal-time quantization coupling by the equal-time expectation value of $\cos(\beta \phi)$ \cite{Burkardt:1992sz,Chen:2023glf,Fitzpatrick:2023aqm}:
\begin{equation}
\lambda_{\rm LC} = \lambda_{\rm ET} \< \cos (\beta \phi)\>. 
\label{eq:LCvsET}
\end{equation}
Even if this expectation value were unknown, it would still be possible to compute all dimensionless ratios of physical quantities in the theory.  In this case, though, the vev is known from integrability, and so we can check the overall scale as well.  We will obtain at high precision the mass spectrum and $c$-function of the model, for all imaginary $\beta$ (sinh-Gordon), and over the range $0 \le \beta \le \sqrt{4\pi}$ for real $\beta$ (sine-Gordon). We compare these quantities to their values computed  from integrability, and find excellent agreement.  We emphasize that this comparison is not simply a quantitative test of the numeric efficiency of the method, but is moreover a stringent test of the following conceptual points:
\begin{itemize}
\item  The relation (\ref{eq:LCvsET}) between the lightcone effective coupling $\lambda_{\rm LC}$, the equal-time coupling $\lambda_{\rm ET}$, and the vev $\< \cos (\beta \phi)\>$.  Existing arguments for this relation include an analysis at all orders in $\beta$ \cite{Burkardt:1992sz}, as well as a matching argument based on integrating out all nonchiral modes to obtain the lightcone effective Hamiltonian, which assumes that only the chiral modes form a complete basis for the lightcone-quantized theory.  
\item The duality of the sinh-Gordon models related by $\beta \rightarrow \frac{8\pi}{\beta}$.    Existing arguments for this duality are based on analytic continuation of the two-to-two scattering amplitude of breathers, but testing it directly has proven to be challenging. In this work, we will compute the $c$-function on both sides of the duality for a range of $\beta$ and find they agree to high precision.   
\end{itemize}

The paper is organized as follows.  First, in section \ref{sec:Computing} we describe a new efficient algorithm for computing the Hamiltonian matrix elements for a generic scalar theory which significantly improves upon an older radial quantization technique.  Then in section \ref{sec:SG} we present the bulk of our numerical results for both the Sine-Gordon and the Sinh-Gordon models, and compare them to integrability.  In section \ref{sec:CFTFormulation}, for the specific case of a single relevant primary deformation of the scalar CFT, we describe the connection between the Fock-space formulation of \ref{sec:Computing}, and the lightcone effective Hamiltonian formulation proposed in \cite{Fitzpatrick:2023aqm}. 
In section \ref{sec:FreeFermion} we demonstrate how bosonization works in the context of the LCT basis, by explicitly demonstrating how at finite truncation, the scalar LCT basis maps to the finite truncation fermionic LCT basis.  In section \ref{sec:Puzzles} we report on some initial efforts to extend the method to less relevant operators, i.e. to $\Delta > 1$.
Finally, we conclude and discuss some potential future research directions in \ref{sec:Future}.

\section{Computing the Hamiltonian}
\label{sec:Computing}
\subsection{Review of Radial Quantization Method}

The Hamiltonians we will consider are sums over terms of the form $\phi^{2n}$, in a basis of states created by (Fourier transforms of) operators of the form $\partial^{k_1} \phi \partial^{k_2} \phi \dots \partial^{k_n} \phi$.  The resulting matrix elements can therefore be computed in principle simply by computing $n$-point functions of $\phi$ in the free theory.  However, in practice, computing such correlators by summing over Wick contractions quickly becomes impractical, due to the rapid growth in the number of possible pairings of operators with $n$.  A much more efficient method to compute the necessary matrix elements is to first construct primary operators $\CO_i$, and compute the matrix elements $\< \CO_i | \phi^{2n}(x) | \CO_j\>$, where $|\CO_i\>$ is the corresponding state in radial quantization.  The speedup in this approach is due to the fact that in radial quantization, the individual factors $\partial^k \phi$ acting on the vacuum create a single Fock space mode, which must either contract with the identical mode (same value of $k$) from the bra state, or else contract with one of the $\phi$s from the interaction $\phi^{2n}$, so that the number of nonvanishing contractions is significantly smaller.  To do the Fourier transforms, the full dependence on the position of the bra and ket operators $\CO_i$ and $\CO_j$ must be restored, which can be done using conformal transformations.  There is one final technical challenge, that $\phi$ is not a local operator in $2d$, so $\phi^{2n}$ does not transform like a local operator. However, one can circumvent this issue by instead calculating $\partial \phi(y_1) \dots \partial \phi(y_n)$, using the fact that $\partial \phi$ {\it is} a primary operator, and then one can integrate over the $y_i$s to get back $\phi(y)$.  This method was first implemented in \cite{Anand:2020gnn}.

\begin{figure}[htbp]
\centering
\includegraphics[width=0.6\linewidth]{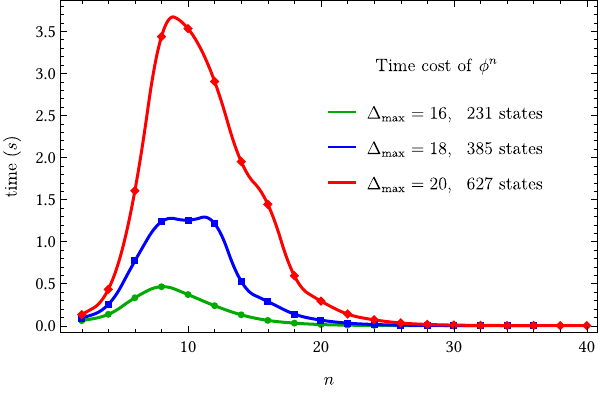}
\caption{\label{fig:benchmarkPhiN}
Benchmark timing of new code computing $\phi^n$ matrix element, for each $n$. The truncation time is peaked at $n=\Dmax/2$. For greater $n$, the states giving nonzero contribution to $\phi^n$ must have large number of particles, and the number of states are suppressed.
}
\end{figure}

The result of this algorithm is that a typical matrix element takes the form
\begin{equation}
\langle \partial^{\kvec}\phi | \phi^{2n} | \partial^{\kvec'}\phi \rangle = \sum_{\substack{\bar \kvec\in \kvec\\\bar \kvec'\in\kvec'}} A_{\kvec,\kvec',\bar\kvec,\bar\kvec'} \mathcal{I}(\bar \kvec, \bar \kvec') \delta_{\kvec / \bar\kvec, \kvec' / \bar\kvec'}
\end{equation}
which consists of the contraction factor $A_{\kvec,\kvec',\bar\kvec,\bar\kvec'}$, which counts all possible ways how radial quantization modes are contracted, and a dynamical part $\mathcal{I}(\bar \kvec, \bar \kvec')$, which encodes the factor from Fourier transforming the CFT correlator. For each possible contraction, there are some number $l(\bar\kvec)$ of $\phi$s that contract to the left, and some number $n-l(\bar\kvec)$ of remaining $\phi$s that contract to the right, so that  $A_{\kvec,\kvec',\bar\kvec,\bar\kvec'}$ factorizes into an ``creation'' part and ``annhilation'' part,
\begin{equation}
A_{\kvec,\kvec',\bar\kvec,\bar\kvec'} = \langle \kvec | \left(\sum_\ell a_\ell^\dagger \right)^{l(\bar\kvec)} | \kvec / \bar\kvec \rangle \langle \kvec / \bar\kvec | \left(\sum_\ell a_\ell \right)^{n-l(\bar\kvec)} | \kvec'\rangle~,
\end{equation}
and moreover each part recursively factorizes into product of ``one-mode annihilation matrix''
\begin{equation}\label{annihilation-recursion}
\langle \kvec | \left(\sum_\ell a_\ell \right)^{m} | \kvec'\rangle
= \sum_{\kvec''} \langle \kvec | \left(\sum_\ell a_\ell \right) | \kvec''\rangle
\langle \kvec'' | \left(\sum_\ell a_\ell \right)^{m-1} | \kvec'\rangle~.
\end{equation}
The key bottleneck is that $A_{\kvec,\kvec',\bar\kvec,\bar\kvec'}$ gets more complicated factorially as $n$ and $\Dmax$ grows. In practice, for $\Dmax=20$ the efficiency is limited by how we store the data. In \cite{Anand:2020gnn}, we stored the information of (\ref{annihilation-recursion}) in two separated matrices: one sparse numerical matrix for coefficients and one symbolical matrix for the list of $k_i$ keeps track of the difference between $\kvec$ and $\kvec'$.  Programming languages such as Mathematica operate on symbolical matrices much more slowly than on numerical matrices, and the main loss of efficiency in the algorithm is solely for this reason.
In this work we have improved the algorithm by storing (\ref{annihilation-recursion}) as a numerical sparse tensor, and the performance is greatly enhanced. A benchmark of the algorithm is shown in Table \ref{tab:benchmarkNewVsOld} and Figure \ref{fig:benchmarkPhiN}. We summarize the improved algorithm in the next subsection.
\begin{table}
\centering
\begin{tabular}{|c|c|c|c|}
\hline
$\Dmax$ & num. of states & code in \cite{Anand:2020gnn} & new code\\
\hline
$20$ & 627 & $3.96s$ & $0.827s$ \\ 
\hline
$40$ & 37338 & $3579s$ & $109s$ \\ 
\hline
\end{tabular}
\caption{\label{tab:benchmarkNewVsOld}
Benchmark timing of new code vs. code in \cite{Anand:2020gnn}, for $\phi^4$ matrix elements computation for all states at $\Dmax = 20$ and $40$. Computation is performed on a Macbook Pro M1.
}
\end{table}

\subsection{Code Implementation and Improved Algorithm}

The new, faster approach adopted in this work is to store the ``annhilation'' factors in one numerical rank-3 tensor. 
In the improved algorithm, the building blocks are generalized ``annihilation operators'' $\,^{[m]}A_{\,\nu; \,\rho}^{\mu}$ keeping track of contracting the interaction with $m$ particles in the ket state
\begin{equation}
\,^{[m]}A_{\,\nu(\kvec); \,\rho(\delta \kvec)}^{\mu(\kvec')} = \begin{cases}
\sqrt{\prod_{k=1}^{\Delta_{\rm max}}(n_k')_{n_k - n'_k} }& {\rm if}~\kvec\backslash\kvec' = \delta \kvec \\ 
0 & {\rm otherwise}
\end{cases}
\end{equation}
where:
\begin{itemize}
\item $\kvec$ is a monomial state $\partial^{k_1} \phi \partial^{k_2} \phi \cdots \partial^{k_{l(\kvec)}} \phi $ and $l(\kvec)$ is the length of the list $\kvec$.
\item Each $\kvec$ has a unique index in the basis of integer partitions of any $K \leq \Delta_{\rm max}$ into exactly $l(\kvec)$ nonzero numbers. We use Greek letters to denote these indices, such as $\mu(\kvec)$. For example, at $\Dmax = 5$ we index all length-3 basis vectors as 
\begin{equation}
\mu(\kvec) : \quad \mu(|111\rangle) = 1,~\mu(|211\rangle) = 2,~\mu(|311\rangle) = 3,~\mu(|221\rangle) = 4~.
\end{equation}
As a special case, for $l(\kvec) = 1$ we have $\mu(\{k\}) = k$.
\item We also give a index $\rho(\delta \kvec)$ for the (un-ordered) list $\delta \kvec$ of the modes that are contracted to the Hamiltonian. The rule of indexing the vectors is the same as above.
\item $\kvec\backslash \kvec'$ is the vector $\kvec$ after removing the subvector $\kvec'$. We additionaly require that $\kvec'\subset \kvec$.
\item $n_k$ is the counting of $k$ in the list $\kvec$.   $(a)_b$  denotes the Pochhammer symbol. 
\end{itemize}
The tensor efficiently encodes all possible ways to get from $|\kvec\rangle$ state to $|\kvec'\rangle$ state by contracting $m$ particle with the Hamiltonian, the symmetry factors, and the levels of all the particles contracted. These tensors of higher $m$ can be recursively constructed from lower $m$. The algorithm for growing one particle $m-1 \rightarrow m$ is a sparse array contraction
\begin{equation}
\,^{[m]}A_{\,\nu; \,\rho}^{\mu} = \,^{[1]}A_{\, \sigma;\, \lambda}^{\mu} \cdot \,^{[m-1]}A_{\,\nu; \,\omega}^{\sigma} \cdot B^{\lambda,\omega}_{\rho}~,
\end{equation}
where we define another rank-3 tensor $B$ to keep track of the contracted modes. 
\begin{equation}
B^{\lambda(\{k\}),\omega(\delta\kvec)}_{\rho(\delta\kvec')} = \begin{cases}
1 & {\rm if}~ \delta\kvec'\backslash\delta\kvec = \{k\}   \\ 
0 & {\rm otherwise}
\end{cases}~.
\end{equation}
The simplest building block of all annihilation operator is the one with $m=1$
\begin{equation}
\,^{[1]}A_{\, \mu(\kvec);\, \nu(\{k\})}^{\rho(\kvec')} = \begin{cases}
\sqrt{n_k} & {\rm if}~\kvec\backslash\kvec' = \{k\} \\ 
0 & {\rm otherwise}
\end{cases}~.
\end{equation} 
which can be efficiently computed because it is highly sparse. The contraction involved in building higher $m$ tensors is also fast as it takes advantage from existing sparse array contraction algorithms that are highly optimized. 

To get the final expression for the matrix element, we take the annhilation operator acting on the out-state $|\kvec'\rangle$ and contract it with the annhilation operator action on the in-state $\langle\kvec|$ and apply the integral formula (\ref{eq:integralformula}) to the contracted $\delta \kvec$ and $\delta \kvec'$
\begin{equation}
\mathcal{M}_{\kvec, \kvec'} = \sum_{m=1}^{n-1} \frac{(4\pi)^{\frac{2-n}{2}}}{m!(n-m)!} \delta_{m,\frac{1}{2}(l(\kvec)-l(\kvec')+n)}~
{(^{[m]}A)^\dagger}_{\,\lambda}^{\mu(\kvec); \,\sigma} \cdot 
\,^{[n-m]} A_{\,\nu(\kvec'); \,\rho}^{\lambda} \cdot C_\sigma^{\rho} \times c_{h, h'}~.
\end{equation} 
The matrix $C$ encodes the result of all spacetime integrals involved in computing the Hamiltonian matrix elements, as follows:
\begin{equation}
\begin{aligned}
F_{\mu(\delta\kvec)}^{\nu(\delta\kvec')}(v) &= \left( \prod_{k\in\delta\kvec} \frac{(1-v^k)}{\sqrt{k}} \prod_{k'\in\delta\kvec'} \frac{(1-v^{-k'})}{\sqrt{k'}} \right) \equiv \sum_s a_s v^s , \\
C_{\mu(\delta\kvec)}^{\nu(\delta\kvec')} &= \left. F_{\mu(\delta\kvec)}^{\nu(\delta\kvec')}(v) \right|_{v^s \rightarrow |s|} = \sum_s a_s |s|,
\end{aligned}
\label{eq:integralformula}
\end{equation}
and the prefactor that depends on the conformal weights of the states is absorbed in $c_{h, h'}$
\begin{equation}
c_{h, h'} = \frac{(-1)^{h-h'}\sqrt{\Gamma(2h)\Gamma(2h')}}{\Gamma(h+h'-1)}~.
\end{equation}
Note that (\ref{eq:integralformula}) we first expand the two products into a single power series in $v^s$, and substitute $v^s \rightarrow |s|$.
This formula was derived in \cite{Anand:2020gnn} and is the result of integrating the interaction $\phi^{2n}(x)$ over the lightcone null surface, as well as Fourier transforming the positions of the bra and ket operators to create momentum-space states.  It takes into account the conformal transformation required to map the bra and ket operators back from $\infty$ and $0$, respectively, to arbitrary spatial positions, as well as the fact that each factor of $\phi$ in $\phi^{2n}$ transforms not like a local operator but rather like the integral of a local operator $\phi \sim \int \partial \phi$.

\section{Application: Free Scalar with Cosine potential}
\label{sec:SG}

\subsection{Sine-Gordon and Sinh-Gordon}
Our main application of our numeric algorithm in this paper is the sine-Gordon theory,\footnote{When $ N_c\equiv \frac{4\pi}{\beta^2}$ is an integer $\ge 2$, the sine-Gordon model also happens to be the IR description of 2d QCD with gauge group $SU(N_c)$ and a Dirac fermion in the fundamental representation in the limit where the bare fermion mass is small \cite{steinhardt1980, Anand:2021qnd}. }
\begin{equation}
	H= \int dx^- \left( \frac{1}{2}(\partial\phi)^2 - \frac{\lambda}{2\pi} \cos(\beta\phi) \right).
	\label{eq:HSG}
	\end{equation}
 It is  integrable for any value of $\beta$, which will allow us to perform explicit checks of all our results.  It can also be  understood purely in CFT terms, as a CFT (a compact free boson, $\phi \cong \phi + \frac{2\pi}{\beta}$) with a relevant deformation.
 The dimension of the $\cos$ operator  is given by 
\begin{equation}
	\Delta_\beta = \frac{\beta^2}{4\pi}.
\end{equation}
The spectrum of the theory contains soliton/anti-soliton and breather bound states, whose masses are given in terms of the coupling constant $\lambda$ by \cite{Zamolodchikov:1995xk} 

\begin{align}
    \textrm{soliton: }& M_s=\left(\frac{\lambda}{2\pi \kappa(\xi)}\right)^{\frac{\xi+1}{2}},\\
	\textrm{breathers: }& m_n=2 M_s \sin \frac{\pi n \xi}{2},\quad n=1,2,...1/\xi,
	\label{eq:SineGordonMasses}
\end{align} 
where $\xi$ and $\kappa$ are given by 
\begin{equation}
	\xi = \frac{\beta^2}{8 \pi-\beta^2} = \frac{\Delta_\beta}{2-\Delta_\beta}, \qquad 
	\kappa(\xi)=\frac{2}{\pi} \frac{\Gamma\left(\frac{\xi}{\xi+1}\right)}{\Gamma\left(\frac{1}{\xi+1}\right)}\left[\frac{\sqrt{\pi} \Gamma\left(\frac{\xi+1}{2}\right)}{2 \Gamma\left(\frac{\xi}{2}\right)}\right]^{2 /(\xi+1)}.
\end{equation}

\begin{figure}[htbp]
\begin{center}
\includegraphics[width=0.5\textwidth]{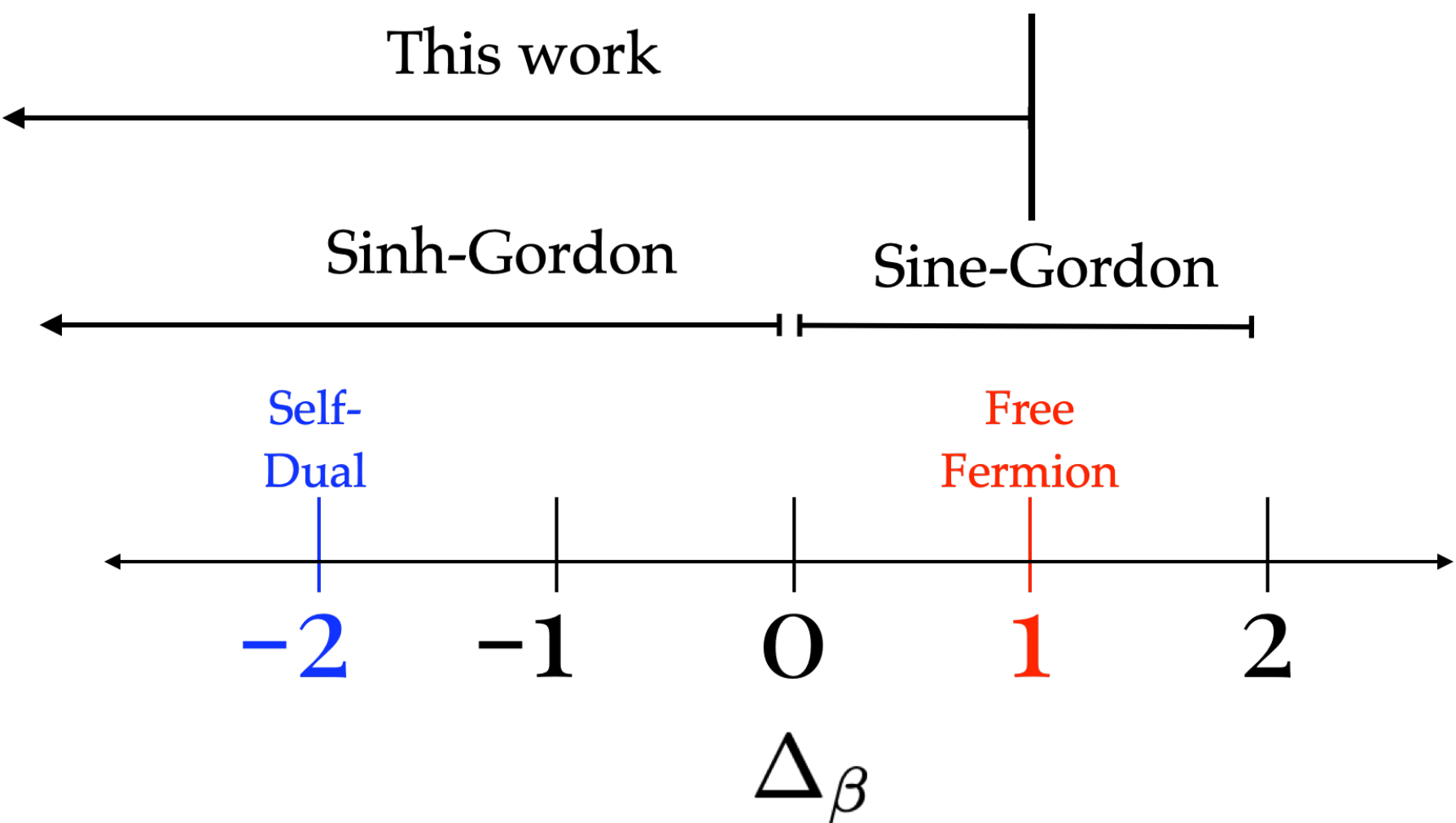}
\caption{Sinh-Gordon and Sine-Gordon parameter space as a function of the scaling dimension  $\Delta_\beta$ of the $V(\phi)$ potential.}
\label{fig:ParameterSpaceCartoon}
\end{center}
\end{figure}

The Sinh-Gordon model is related to the Sine-Gordon model by analytic continuation, by taking $\beta$  to be pure imaginary.  In that case, the parameter $\xi$ becomes negative, indicating that the breathers disappear and solitons do not have any stable bound states. We will mostly use $\Delta_\beta$ to parameterize the space of theories, in order to emphasize the intuition from thinking of  the theory as a CFT deformed by a relevant operator.  In terms of $\Delta_\beta$, the (conjectured) duality is
\begin{equation}
\Delta_\beta \leftrightarrow \frac{4}{\Delta_\beta},
\end{equation}
and the self-dual point is $\Delta_\beta=-2$.  The sinh-Gordon regime is $\Delta_\beta <0$ whereas the sine-Gordon regime is $2> \Delta_\beta >0$.  The point $\Delta_\beta=1$ is dual to a free Dirac fermion. See Fig.~\ref{fig:ParameterSpaceCartoon} for a cartoon summary of the parameter space.

\subsection{Mass Spectrum}

We evaluate the matrix elements of $\cos (\beta \phi)$ by Taylor expanding 
\begin{equation}
\cos ( \beta \phi) = \sum_{n=0}^\infty \frac{ (-1)^n(\beta \phi)^{2n}}{(2n)!},
\end{equation}
and evaluating term-by-term. For any individual matrix element between a fixed pair of basis states, only a finite number of terms in the Taylor series will contribute.  This is because if the power of $\phi^{2n}$ is too large, there are simply not enough $\phi$s in the bra and ket states to contract with, so the contribution vanishes. Therefore, the sum over $n$ is manifestly convergent.\footnote{Using Wick contractions of $\phi$s is also equivalent to using the Fock space basis for free massless scalars, as was done in \cite{Burkardt:1992sz} in a DLCQ basis.
% The paper \cite{Burkardt:1992sz} also gave an argument that in the sine-Gordon model, the effect of lightcone zero modes was simply to renormalize the coefficient of the cosine potential. 
We will be able to improve on the results of \cite{Burkardt:1992sz} by going to much higher truncations, and also by using an LCT basis which has faster convergence. }

Perhaps the simplest observable one can consider is the mass gap of the theory.  There is one obstacle to comparing our numeric results for the gap to the analytic formulas in the previous subsection, because one has to match the parameters in lightcone quantization to the corresponding parameters in equal-time quantization, where the analytic expressions were derived.  This matching normally must be done numerically \cite{Chen:2023glf}.  In this case, the relation between the parameters is
\begin{equation}
\lambda_{\rm LC} = \lambda_{\rm ET} \< \cos \beta \phi\>_{\rm ET},
\end{equation}
where $\lambda_{\rm ET}/2\pi$ is the coefficient of $\cos \beta \phi$ in the equal-time Lagrangian, and $\< \cos \beta \phi\>_{\rm ET}$ is the vacuum expectation value in equal-time.\footnote{This relation was first derived in \cite{Burkardt:1992sz}. It was also derived in a more general CFT framework in \cite{Chen:2023glf,Fitzpatrick:2023aqm}.  One subtlety is that in the general CFT framework, the lightcone interaction for a relevant deformation $\CO_R$ takes the nonlocal form $\CO_R(x)\CO_R(\infty)$, whereas the $\cos$ potential in the Fock space prescription is apparently local; in section \ref{sec:CFTFormulation}, we explain how these two apparently different prescriptions end up being identical. }

Fortunately, exactly this expectation value was derived analytically in \cite{Konik:2020gdi}, where they found the elegant relation
\begin{equation}
 \lambda_{\rm ET} \< \cos \beta \phi\>_{\rm ET}   = M_s^2 \frac{\pi \tan \left( \frac{\pi \Delta_\beta}{4-2\Delta_\beta} \right)}{2-\Delta_\beta},
 \label{eq:VevReln}
\end{equation}
where $M_s$ is the mass of the soliton, see (\ref{eq:SineGordonMasses}).  Therefore, when we use the resulting relation for $\lambda_{\rm LC}$ in the lightcone effective action, we automatically obtain the spectrum of mass-squareds as a multiple of $M_s^2$.

 The fact that the coefficient $\lambda_{\rm LC} =  \lambda_{\rm ET} \< \cos \beta \phi\>_{\rm ET}$ from (\ref{eq:VevReln}) is linear in $M_s^2$ rather than in $M_s$ is in fact a nontrivial requirement of the simple existence of the lightcone Hamiltonian, which computes the spectrum of $P^2$ directly.\footnote{The key point is that there are no other kinematic scales that could compensate the difference in units between $M_s$ and $M_s^2$, because lightcone is manifestly momentum- (by Lorentz invariance of the null quantization surface) and volume-independent (because of infinite volume). }  
One interesting feature that we will return to is that for a fixed equal-time coefficient $\lambda_{\rm ET}$, the vev $\< \cos \beta \phi\>$ diverges like $\frac{1}{\Delta_\beta-1}$ as $\Delta_{\beta} \sim 1$, so in order for lightcone quantization to correctly reproduce the finite mass spectrum  (which it does) it must have an infinite number of states that become massless if the lightcone coupling $\lambda_{\rm LC}$ is held fixed as $\Delta_\beta \rightarrow 1$.

\begin{figure}[htbp]
\begin{center}
\includegraphics[width=0.49\textwidth]{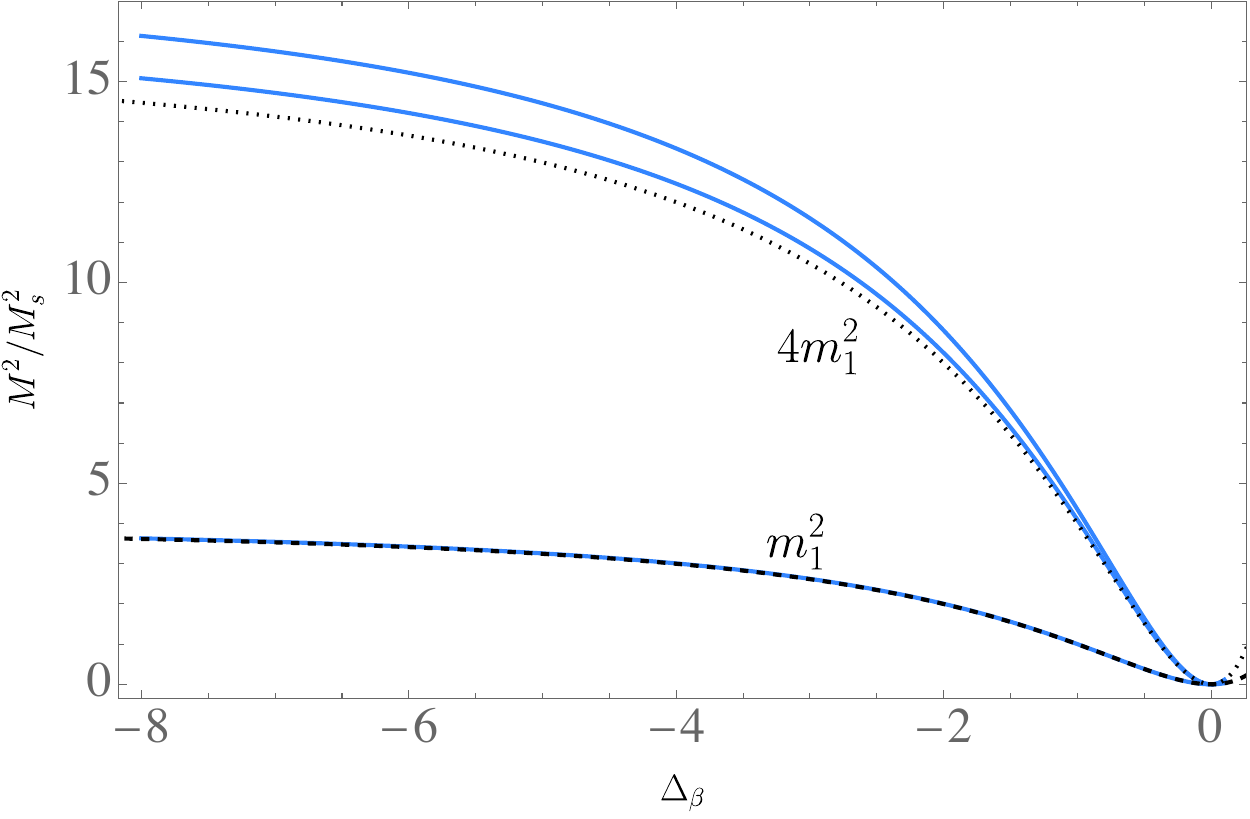}
\includegraphics[width=0.492\textwidth]{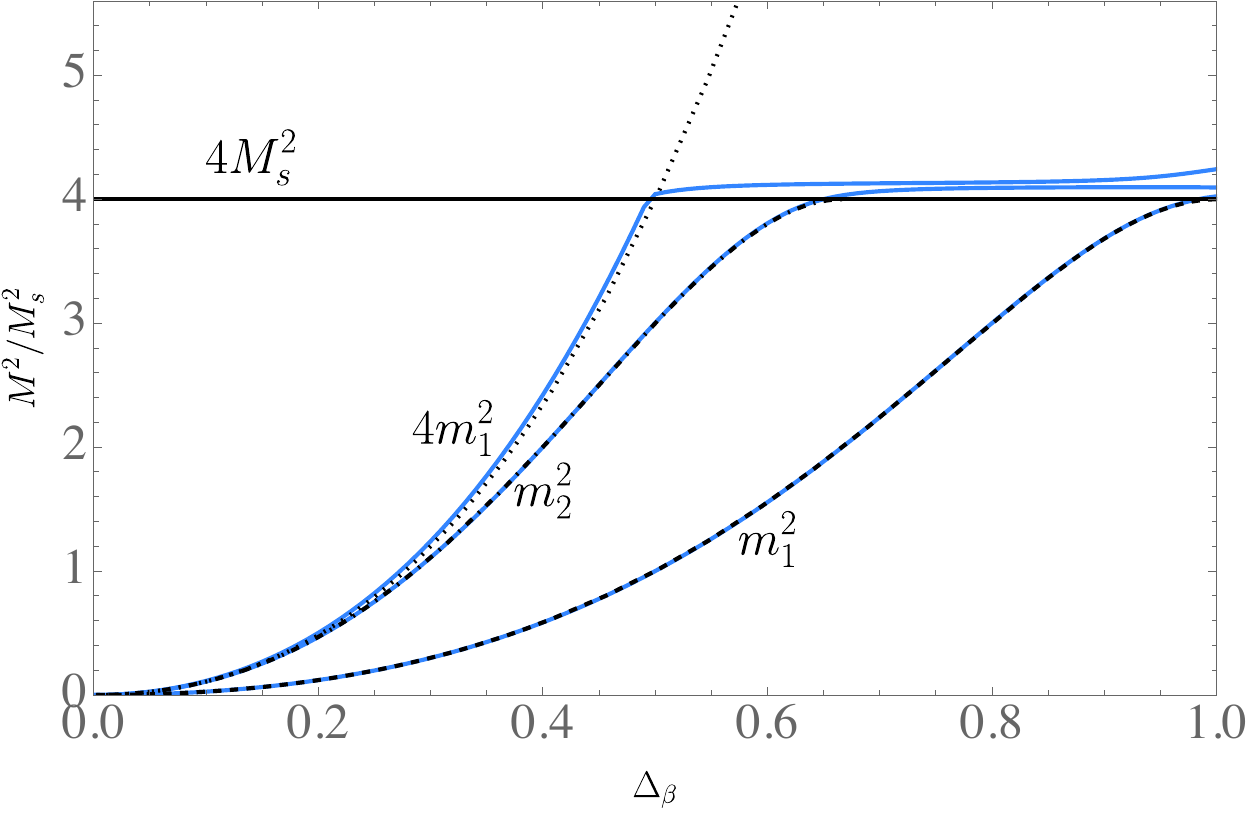}
\includegraphics[width=0.485\textwidth]{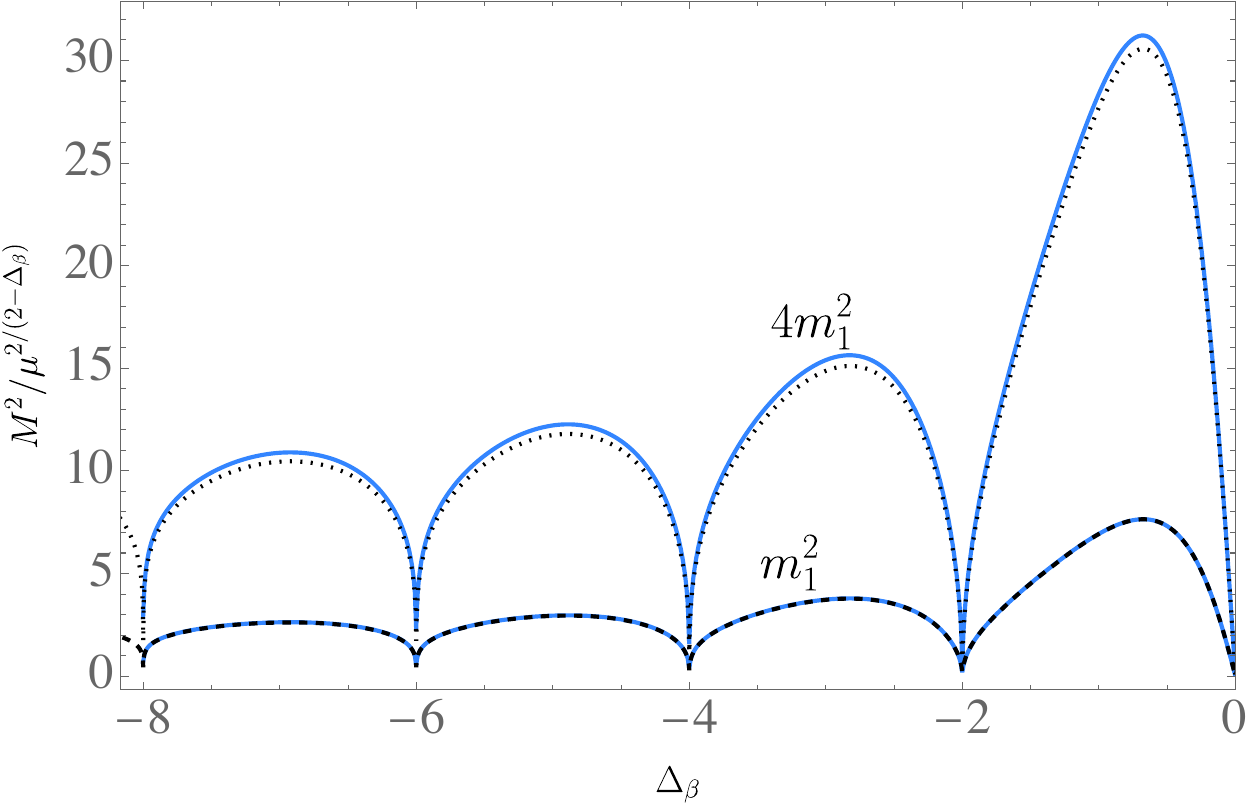}
\includegraphics[width=0.501\textwidth]{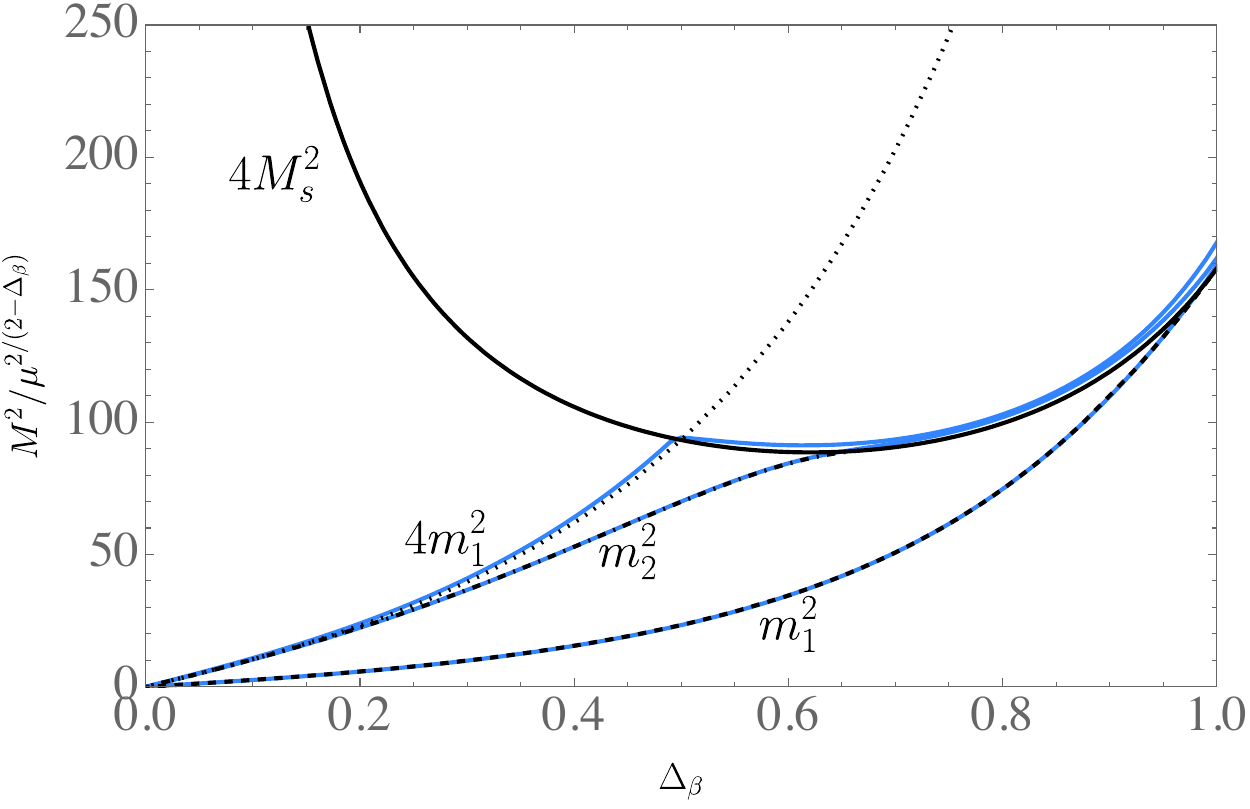}
\caption{Comparison of low-lying states in LCT versus analytic prediction.  The top two plots show the spectrum of mass-squareds divided by the soliton mass-squared $M_1^2$, for sinh-Gordon ($\Delta_{\beta}<0$, {\it left}) and sine-Gordon ($\Delta_\beta>0$, {\it right}).  The lightest three states from truncation are shown in light blue, solid.  We also show the theoretical prediction from integrability methods for the first breather $m_1^2$ ({\it black, dashed}), second breather $m_2^2$ ({\it black, dot-dashed}), and the two-particle thresholds for the multi-breather $4m_1^2$ ({\it black, dotted}) and multi-soliton $4M_s^2$ states   ({\it black, solid}), in the range where these states are physical (the single-soliton state is charged and therefore does not show up as an eigenvalue in the neutral sector we are considering).  In the bottom two plots we show the same spectral results but in units of the equal-time coupling, $2\mu \equiv \lambda_{\rm ET}/2\pi$. The LCT data is computed under truncation $\Dmax = 20$.
}
\label{fig:SpectrumComparison}
\end{center}
\end{figure}

 In Fig.~\ref{fig:SpectrumComparison}, we show our spectrum from LCT over the range $-8 < \Delta_\beta < 1$, and compare to integrability results.  
Single soliton states are not visible in the spectrum, since they carry solitonic charge and we only evaluate the neutral sector.   The second breather is part of the spectrum only for $0 < \Delta_\beta < 2/3$, and disappears into the continuum of multi-soliton states for larger values of $\Delta_\beta$.  The masses of the remaining breathers ($n\ge 3$) are either above $2m_1$, or else absent from the spectrum.

\begin{figure}
\centering
\includegraphics[width=0.4\linewidth]{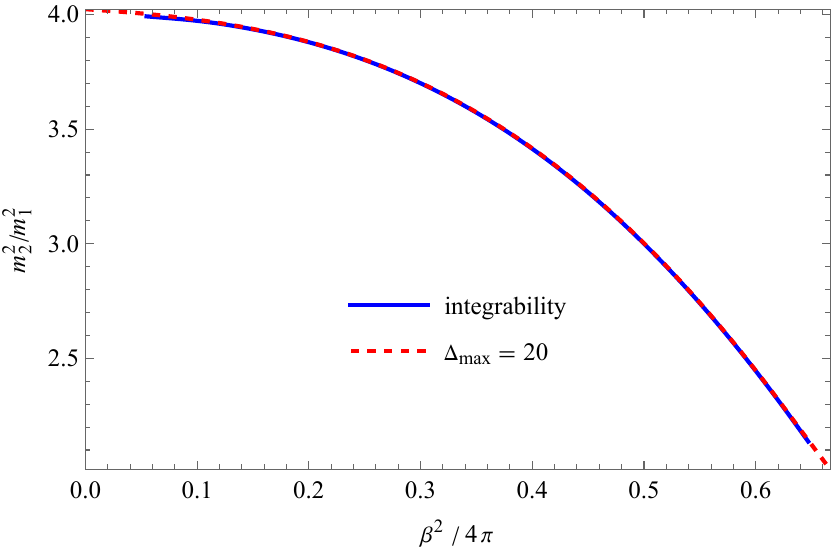}
\includegraphics[width=0.4\linewidth]{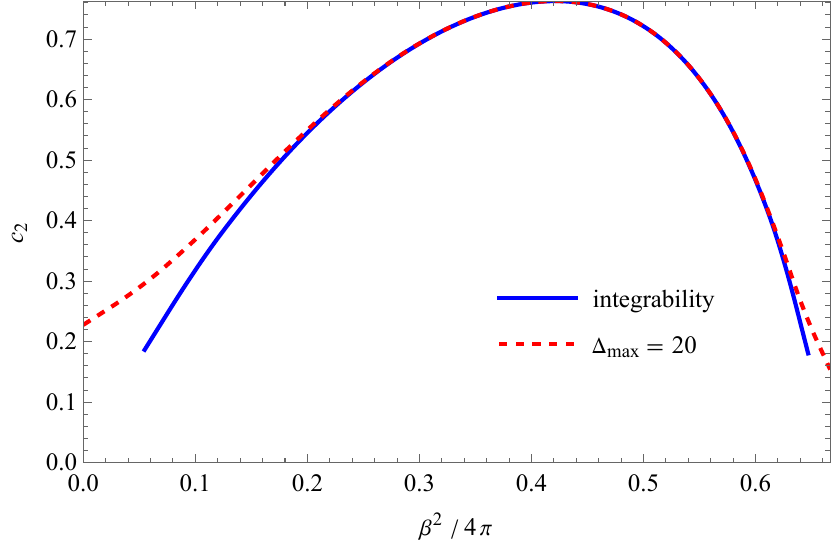}
\caption{
 Sine-Gordon from LCT scalar code. {\it Left}:   mass ratio $m_2^2/m_1^2$ as a function of $\Delta_\beta = \frac{\beta^2}{4\pi}$ from LCT vs. the integrability result. {\it Right}: even bound state $c_2$ spectral density. 
} 
\label{fig:LCTSineGordon}
\end{figure}

For the sine-Gordon model,  we also evaluate the one-particle form factor $c_2$ for the stress tensor to create the second breather; equivalently, $c_2$ is the contribution to the  $c$-function from the second breather.\footnote{In this case the corresponding one-particle form factor  $c_1$ for the first breather vanishes identically because there is a $\mathbb{Z}_2$ symmetry ($\phi \rightarrow -\phi$) under which $T_{--}$ is even but the first breather is odd. }
   From (\ref{eq:SineGordonMasses}), the second breathers is physical when $\Delta_\beta<2/3$.  In Figure \ref{fig:LCTSineGordon}, we compare the value of $c_2$, as well as the dimensionless ratio $m_2^2/m_1^2$, from truncation against the predicted value from integrability, for the full range $0 < \Delta_\beta < \frac{2}{3}$, and find excellent agreement. The main source of error in the calculation of $c_2$ arises near the boundaries of this range, due to mixing with the multiparticle states (multi-breather states near $\Delta_\beta \sim 0$ and multi-soliton states at $\Delta_{\beta} = 2/3$).

\subsection{\texorpdfstring{$c$}{c}-function  of Sine-Gordon}

 To perform more stringent tests, we can  look at more complicated observables, and the main one we will consider is the Zamolodchikov $c$-function, defined as an integral of the $T_{--} = (\partial_- \phi)^2$ two-point function spectral density $\rho_{T_{--}}$:
\begin{equation}
c(s) \equiv \frac{12\pi}{p_-^4} \int_0^{s} d\mu^2  \rho_{T_{--}}(\mu^2), \quad 
\rho_{T_{--}}(\mu^2) = \big| \langle (\partial_- \phi)^2 | \mu, p \rangle \big|^2 
\label{eq:cfuncDef}
\end{equation}

\begin{figure}[ht]
\begin{center}
\includegraphics[width=0.4\linewidth]{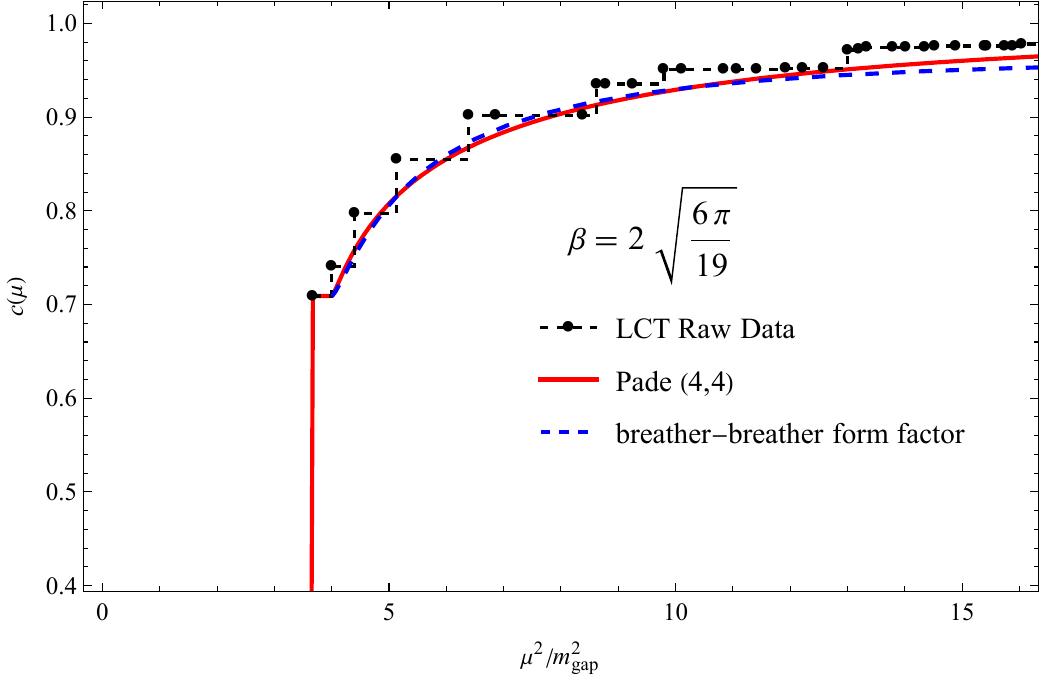}
\includegraphics[width=0.4\linewidth]{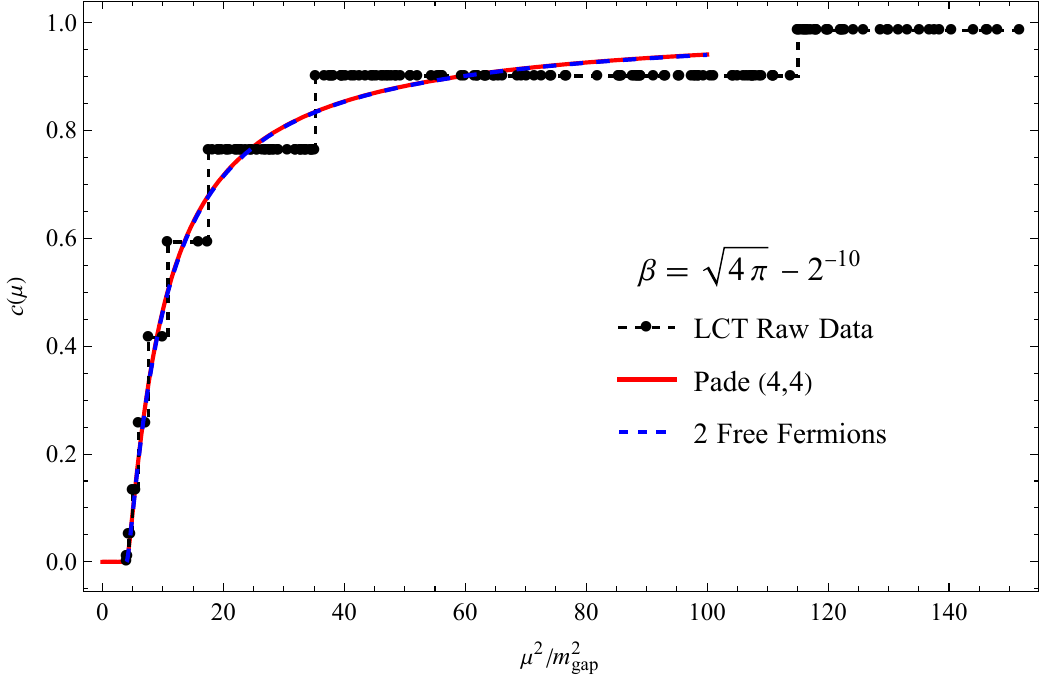}
\caption{{\it Left:}  $c$-function at an (ad hoc) intermediate value of $\beta$ ($\Delta_\beta=0.316$) from LCT compared with the integrability result counting the lowest parity even bound state and the breather-breather two-particle form factor.  {\it Right:} The $c$-function at $\beta$ infinitesimally close to $\sqrt{4\pi}$, compared to the free complex fermion.  We have also improved the `raw' $c$-function from LCT, which has visible discrete jumps due to the discrete nature of the spectrum in any finite truncation, into a smooth function (labeled ``Pade (4,4)'') using the Pad\'e approximation technique described in the text. The LCT data is computed under truncation $\Dmax=20$. }
\label{fig:SineGordonCFunc}
\end{center}
\end{figure}

In Fig.\ref{fig:SineGordonCFunc}, we show this $c$-function computed from LCT at two different values of $\beta$, one at an intermediate value (of no particular special significance) and another extremely close to the point $\beta=\sqrt{4\pi}$ ($\Delta_\beta=1$) where the theory is dual to a free fermion.\footnote{The reason we move slightly $(2^{-10}$) below the free fermion point is that exactly at the free fermion point, the Hamiltonian contains a ratio of the form $0/0$, and in practice we obtain $\beta = \sqrt{4\pi}$ by taking a limit, as we discuss in detail in section \ref{sec:FreeFermion}.}  In the former case, we compare the result to the $c$-function from integrability, whereas in the latter case we can compare to the simple closed form of the $c$ function for a free fermion:
\begin{equation}
c_{\textrm{ free fermion}}(s) = \frac{1}{2} \left( 1 - \frac{4 M_s^2}{s}\right)^{1/2}.
\end{equation}
Because the spectrum we obtain from truncation is discrete (due to the fact that we are always keeping only a finite-dimensional subspace of the full Hilbert space), the $c$-function we obtain from truncation is a sum over step functions, as one can see from the definition (\ref{eq:cfuncDef}) and the fact that the spectral density $\rho_{T_{--}}$ is a sum over $\delta$ functions.  We refer to this sum over step functions as the `raw' LCT spectral density.  We can improve it to a smooth function using the following method from \cite{Chen:2021bmm}.  First, consider the spectral representation of the time-ordered two-point function of $T_{--}$:
\begin{equation}
\Delta_{T_{--}}(s) =i  \int_0^\infty  d\mu^2 \frac{\rho_{T_{--}}(\mu^2)}{s-\mu^2 + i \epsilon}.
\end{equation}
We compute all of the Taylor series coefficients of $\Delta_{T_{--}}(s)$ in an expansion in $s$ around 0:
\begin{equation}
\Delta_{T_{--}}(s)= -i  \frac{p_-^4}{12\pi} \sum_{n=0}^\infty  \< \mu^{-2n} \> s^n, \quad \< \mu^{-2n} \> \equiv   \int_0^\infty d\mu^2 \frac{c'(\mu^2)}{\mu^{2n}} = \int_0^{\infty} d\mu^2 \mu^{-2n} \frac{12\pi}{p_-^4} \rho_{T_{--}}(\mu^2)  .
\label{eq:MomentDefn}
\end{equation}
The time-ordered two-point function has a branch cut at the two-particle threshold $s=4m^2$.\footnote{Any poles below this threshold show up as explicit stable states in the spectrum obtained numerically, and can therefore be removed by hand.}  We define the following variable $\rho$ which maps this branch cut to the unit circle in the complex plane:
\begin{equation}
s = \frac{16 m^2 \rho}{(1+\rho)^2}.
\end{equation}
Then, we use the series expansion in $s$ to determine the series expansion in $\rho$, which we use to construct the Pad\'e approximant to the time-ordered two-point function.  Finally, we can take the imaginary part of this Pad\'e approximant to recover the spectral density $\rho_{T_{--}}$, but now in the form of a continuous function.\footnote{The exact result depends one which Pad\'e order one chooses, but as long as the truncation is high enough and many terms $a_n$ in the expansion have converged, then there is a window of different Pad\'e orders over which the result is nearly independent of the order.  In practice, we have chosen to use the $(4,4)$ approximant in the $\rho$ variable.}

An important advantage of the $c$-function, compared to the spectrum alone, is that it contains detailed information about the multi-particle spectrum.  Such information becomes more important for $\Delta_\beta$ outside the range $0< \Delta_\beta < 2/3$, where there are fewer dimensionless quantities to check because there is only one stable breather mode.  In particular, for the sinh-Gordon model, where there is only one stable one-particle state, there are no dimensionless ratios of particle masses and one is forced to consider multi-particle observables in order to analyze the theory.  We now turn to such an analysis in the next subsection.

\subsection{\texorpdfstring{$c$}{c}-function of Sinh-Gordon and Duality}

To test the duality $\Delta_\beta \leftrightarrow 4/\Delta_\beta$ of the sinh-Gordon model, we need to compare physical observables involving dimensionless ratios.  We will begin by choosing units where the breather mass is $m_1=1$.  For any value of $\Delta_\beta <0$, the spectrum is then simply an isolated state at $\mu^2=1$, followed by a multiparticle continuum starting at $\mu^2 \ge 4$, and so is not sufficiently rich to provide compelling evidence for or against the existence of the duality.  Instead, the observable we will focus on in this subsection is the Zamolodchikov $c$-function, defined in (\ref{eq:cfuncDef}).  As discussed in the previous subsection, we get the most accurate calculation of the $c$ function by computing moments of $\mu^{-2n}$ integrated against the spectral density, as defined in (\ref{eq:MomentDefn}). 

\begin{figure}[ht]
\begin{center}
\includegraphics[width=0.8\textwidth]{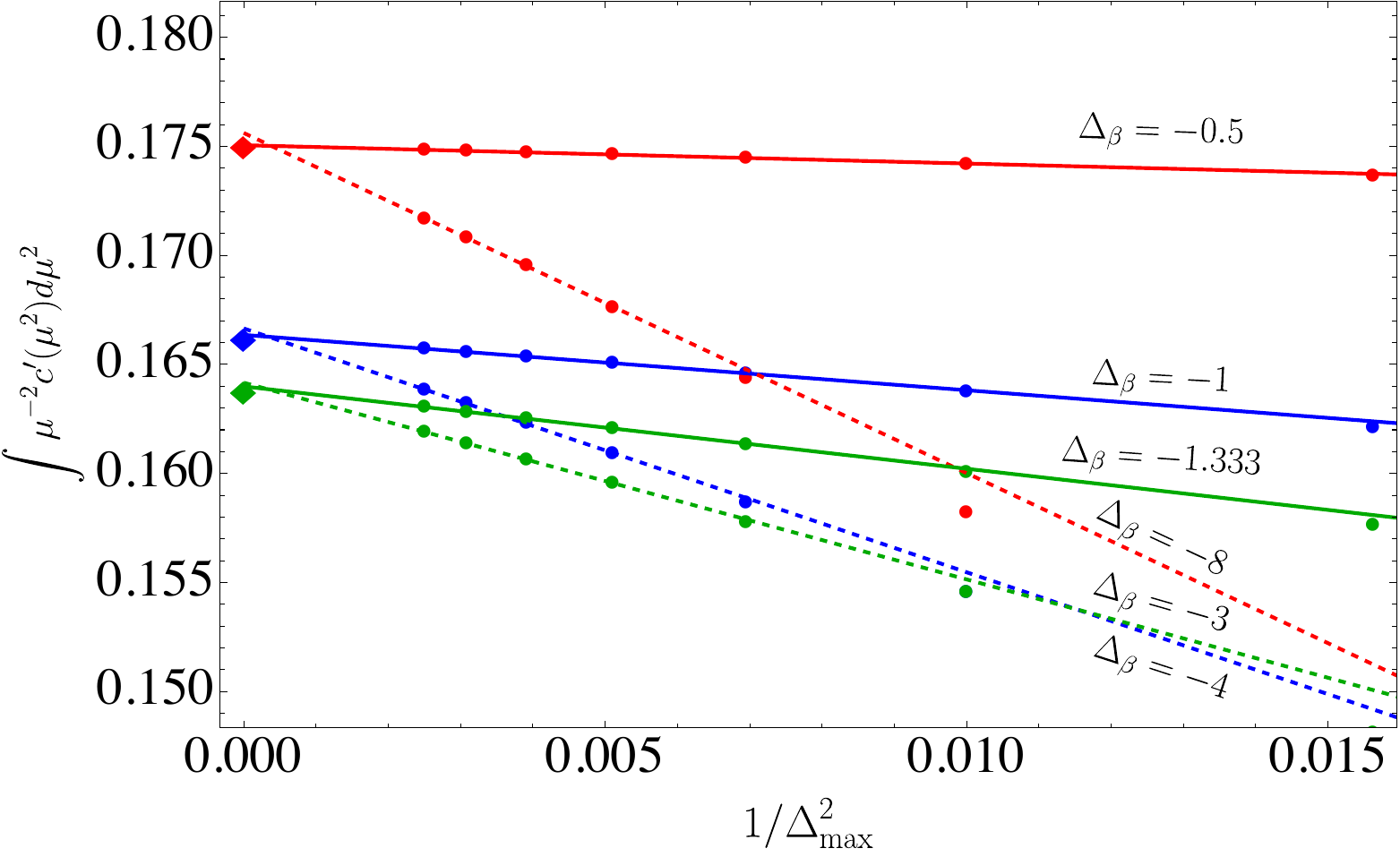}
\caption{Extrapolating the moment $\< \mu^{-2}\>$ as a function of truncation cutoff $\Delta_{\rm max}$ for several values of $\Delta_\beta$. The truncation results are shown as circles, and the results from integrability are shown as diamonds. The lines are linear fits $a+b/\Delta_{\rm max}^2$ performed by using the largest 4 values of $\Delta_{\rm max}$ ($\Delta_{\rm max} = 14,16,18,20$).  Values of $\Delta_\beta$ related by duality $\Delta_\beta \leftrightarrow 4/\Delta_\beta$ are shown using the same color.}
\label{fig:SGExtrapolating}
\end{center}
\end{figure}

Another practical advantage of computing moments $\< \mu^{-2n}\>$ rather than the $c$-function directly is that it is straightforward to extrapolate the moments as a function of the truncation parameter $\Delta_{\rm max}$. In Fig. \ref{fig:SGExtrapolating}, we show how we do this for the leading nontrivial moment $\< \mu^{-2}\>$,\footnote{The moment $\< \mu^0\>$ is just the UV central charge, which is exactly equal to 1 for any point in parameter space.  Moreover, it is exactly equal to 1 in our numeric Hamiltonian truncation calculation, for any value of the truncation $\Delta_{\rm max}$, because it is simply given by the normalization of the basis state created by the energy-momentum tensor. } for several values of $\Delta_\beta$.  Once can see that at the largest values of $\Delta_{\rm max}$ that we used, $\Delta_{\rm max}=20$, the moments are already quite accurate, and nearly the same in models related by the duality.  However, the agreement improves even more if we extrapolate, by fitting the last four values of $\Delta_{\rm max}$ (i.e. $\Delta_{\rm max}=14,16,18,20)$ to $a+b/\Delta_{\rm max}^2$.  The convergence is generally faster for the value of $\Delta_\beta$ closer to 0 (i.e. the fit parameter $b$ is smaller).  Finally, we also show for comparison the prediction for $\<\mu^{-2}\>$ from integrability, and find excellent agreement. 

\begin{figure}[ht]
\begin{center}
\includegraphics[width=1.\textwidth]{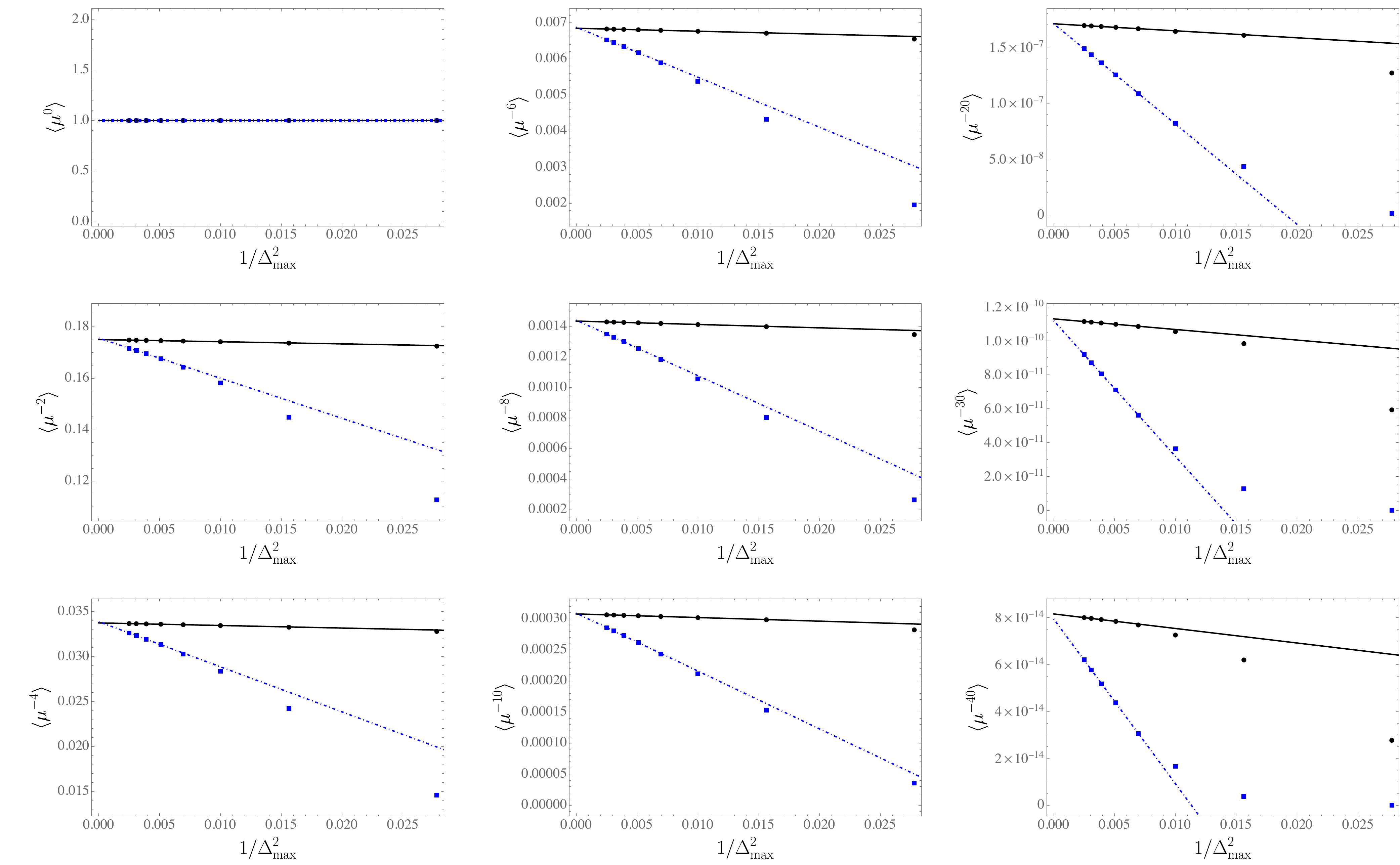}
\caption{Extrapolating the moments $\int d\mu^2 \mu^{-2n} c'(\mu^2)$ for the Zamolodchikov $c$-function $c(\mu^2)$ as a function of the truncation $\Delta_{\rm max}$, for $\Delta_\beta=-1/2$ ({\it black squares}) and $\Delta_{\beta}=-8$ ({\it blue circles}), which are related by the duality of the sinh-Gordon model.  The largest four values of $\Delta_{\rm max}$ are used to perform a linear extrapolation to $\Delta_{\rm max}=\infty$ ({\it black solid} at $\Delta_{\beta}=-1/2$, {\it blue dot-dashed} at $\Delta_{\beta} =-8$). The moment $n=0$ is exactly the UV central charge $c=1$ for any $\Delta_{\rm max}$ and any $\Delta_{\beta}$. }
\label{fig:SGExtrapMany}
\end{center}
\end{figure}

To fully check the agreement of the $c$-function, we should compare as many of its moments as possible.  In practice, even just the first 10 or so moments are sufficient to construct Pad\`e approximants that accurately reproduce the full $\mu$-dependence of the $c$-function, but the convergence of our method is fast enough that we can compute many more moments than needed.  In Fig. \ref{fig:SGExtrapMany}, we consider the dual points $\Delta_\beta=-1/2$ and $\Delta_\beta=-8$, and show the extrapolation for many moments, from $\< \mu^{-2}\>$ up to $\< \mu^{-40}\>$, and find that the $\Delta_{\rm max} \rightarrow \infty$ values agree to high precision; the numeric values are shown for comparison in Table \ref{tab:Moments} for $2n=2, 4, \dots, 40$.

\begin{table}
\begin{center}
\begin{tabular}{l c c c}
\hline
$2n$ & $\Delta_\beta=-1/2$ & $\Delta_\beta=-8$ & integrability \\
\hline
2 & 0.17505107 & 0.17561363 & 0.17483882 \\
4 & 0.03374801 & 0.03388418 & 0.03374143 \\
6 & 0.00684630 & 0.00687572 & 0.00684569 \\
8 & 0.00143513 & 0.00144139 & 0.00143497 \\
10 & 0.00030789 & 0.00030920 & 0.00030784 \\
12 & 0.00006721 & 0.00006748 & 0.00006720 \\
14 & 0.00001487 & 0.00001492 & 0.00001487 \\
16 & 3.328 $ \times 10^{-6} $ & 3.336 $ \times 10^{-6} $ & 3.326 $ \times 10^{-6}$ \\
18 & 7.512 $ \times 10^{-7} $ & 7.525 $ \times 10^{-7} $ & 7.508 $ \times 10^{-7}$ \\
20 & 1.709 $ \times 10^{-7} $ & 1.710 $ \times 10^{-7} $ & 1.708 $ \times 10^{-7} $\\
22 & 3.912 $ \times 10^{-8} $ & 3.908 $ \times 10^{-8} $ & 3.909 $ \times 10^{-8} $\\
24 & 9.007 $ \times 10^{-9} $ & 8.982 $ \times 10^{-9} $ & 8.999 $ \times 10^{-9}$ \\
26 & 2.084 $ \times 10^{-9} $ & 2.074 $ \times 10^{-9} $ & 2.082 $ \times 10^{-9}$ \\
28 & 4.842 $ \times 10^{-10} $ & 4.807 $ \times 10^{-10} $ & 4.836 $ \times 10^{-10} $\\
30 & 1.129 $ \times 10^{-10} $ & 1.118 $ \times 10^{-10} $ & 1.128 $ \times 10^{-10}$ \\
32 & 2.643 $ \times 10^{-11} $ & 2.610 $ \times 10^{-11} $ & 2.639 $ \times 10^{-11}$ \\
34 & 6.206 $ \times 10^{-12} $ & 6.108 $ \times 10^{-12} $ & 6.195 $ \times 10^{-12}$ \\
36 & 1.461 $ \times 10^{-12} $ & 1.433 $ \times 10^{-12} $ & 1.458 $ \times 10^{-12}$ \\
38 & 3.449 $ \times 10^{-13} $ & 3.371 $ \times 10^{-13} $ & 3.441 $ \times 10^{-13}$ \\
40 & 8.161 $ \times 10^{-14} $ & 7.946 $ \times 10^{-14} $ & 8.140 $ \times 10^{-14}$ \\
\hline
\end{tabular}
\end{center}
\caption{Values of the moments $\< \mu^{-2n}\>$ at $\Delta_\beta=-1/2,-8$, compared from their extrapolated values at $\Delta_{\rm max}=\infty$ in LCT (first two columns) and from integrability (last column). }
\label{tab:Moments}
\end{table}

\begin{figure}[ht]
\begin{center}
\includegraphics[width=1.\textwidth]{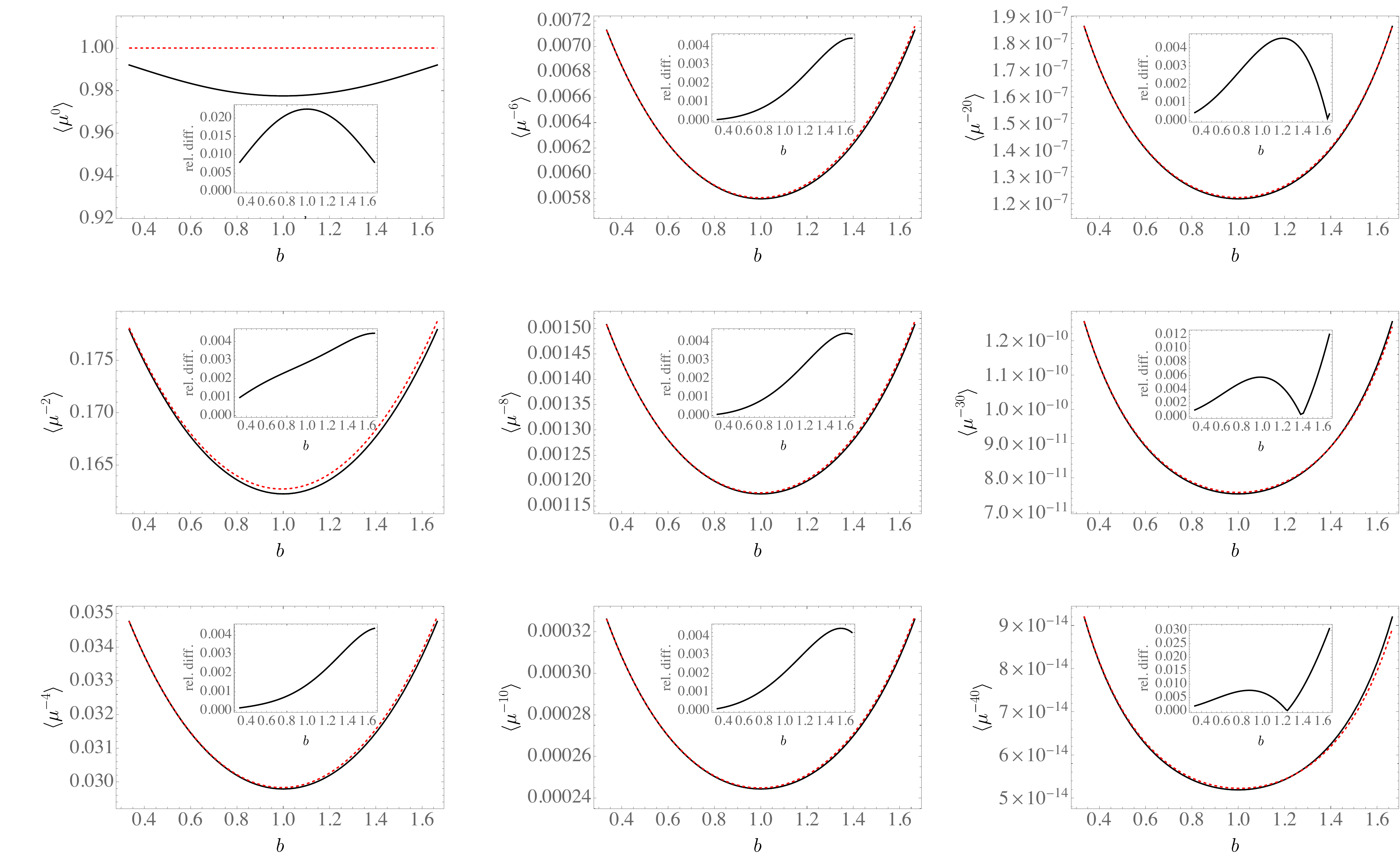}
\caption{Moments $\int d\mu^2 \mu^{-2n} c'(\mu^2)$ for the Zamolodchikov $c$-function $c(\mu^2)$, from LCT ({\it red, dashed}) versus integrability ({\it black, solid}), as a function of  $b \equiv \frac{2\beta^2}{\beta^2-8\pi} =-\frac{2\Delta_\beta}{2-\Delta_\beta}$.   The integrability result is exactly invariant under the duality $b \leftrightarrow 2-b$, so the fact that LCT result closely approximates the integrability results on both sides of $b=1$ is a quantitative test of the duality.  The inset shows the relative difference $\equiv \left| \frac{\<\mu^{-2n}\>_{\rm int} - \<\mu^{-2n}\>_{\rm LCT}}{\<\mu^{-2n}\>_{\rm int}}\right|$. The $n=0$ moment is exactly the central charge $c=1$ of the UV, which is exactly reproduced by LCT but only approximately reproduced by the integrability result, since the latter only includes contributions from two-particle states. }
\label{fig:SGMoments}
\end{center}
\end{figure}

Next, in Fig. \ref{fig:SGMoments}, we show these moments (extrapolated to $\Delta_{\rm max}=\infty$) for a continuous range of parameters of the model.  To make the duality more intuitively apparent by eye, we plot the moments as a function of the parameter 
\begin{equation}
b \equiv - \frac{2\Delta_\beta}{2-\Delta_\beta},
\end{equation}
since under the duality $b \leftrightarrow 2-b$.  We also show a prediction from integrability for comparison.  The integrability prediction is only an approximation to the exact result, since it only includes contributions to the $c$-function from two-particle states.  Because the moments $\mu^{-2n}$ are IR-dominated, and increasingly so at larger $n$, the integrability result is a decent approximation at small $n$ (less than 2\% error for $n=0$) and rapidly becomes an excellent approximation at larger $n$.  Moreover, this approximation from integrability is exactly invariant under the duality, so its agreement with the LCT result on both sides of the duality in Fig.  \ref{fig:SGMoments} provides a simple quantitative visualization of the test of duality with our truncation methods.\footnote{The fact that the $c$-function is accurate all the way up to the self-dual point, and past it, is a significant advantage of lightcone quantization over TCSA (i.e. equal-time quantization). By contrast, in TCSA\cite{Konik:2020gdi}, convergence appears to fail as one approaches the self-dual point. The $c$-function has not been computed in TCSA.
%.  By contrast, in TCSA, convergence appears to fail as one approaches the self-dual point.
%the sinh-Gordon model has the additional challenge of dealing with the zero mode of $\phi$, {\red say more, cite Konik, reread his paper to make sure the problem really is the zero mode}, which becomes more serious the closer one is to the self-dual point.
}

\begin{figure}[htbp]
\begin{center}
\includegraphics[width=0.4\textwidth]{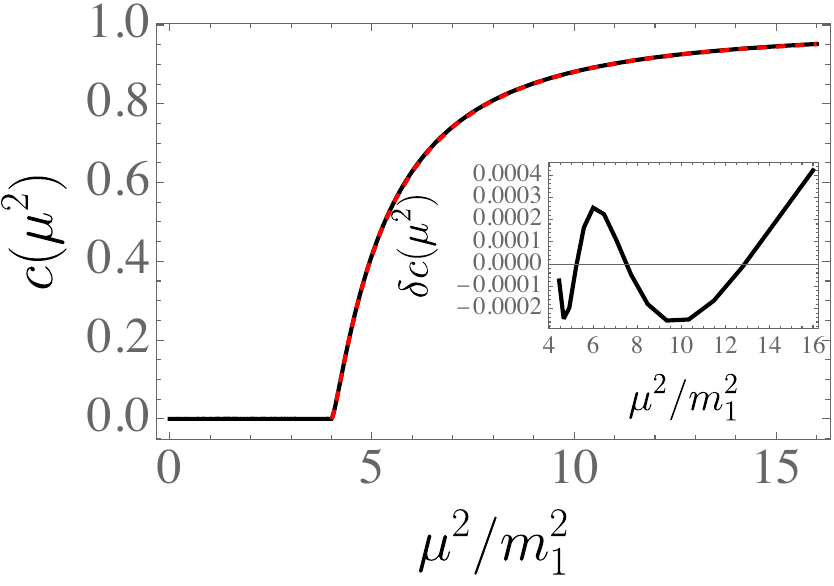}
\includegraphics[width=0.4\textwidth]{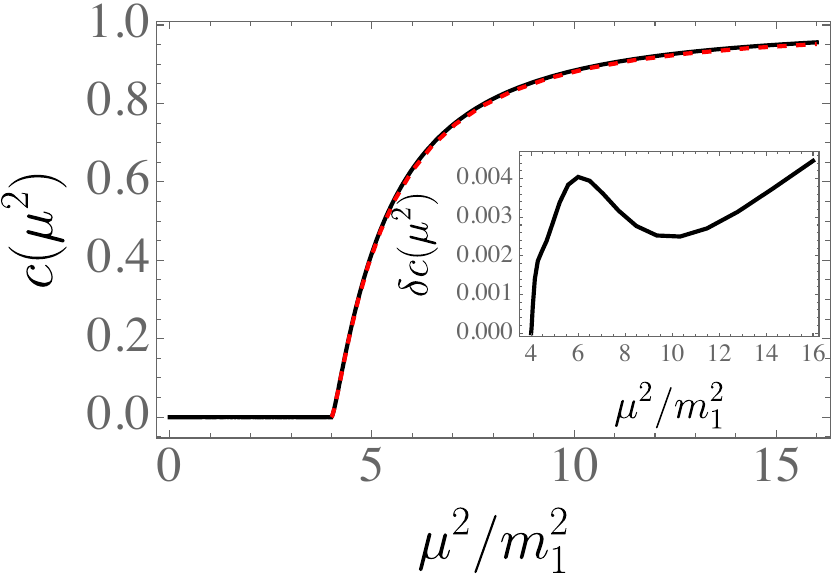}
\caption{Zamolodchikov $c$-function at $\Delta_\beta=-1/2$ ({\it left}) and $\Delta_\beta=-8$ ({\it right}), from LCT with Pad\'e improvement ({\it black, solid}) and integrability ({\it red, dashed}). Inset shows the difference between the two, $\delta c(\mu^2) \equiv c_{\rm trunc}(\mu^2) - c_{\rm integ}(\mu^2)$. }
\label{fig:cFuncSGExample}
\end{center}
\end{figure}

Finally, to emphasize that the accurate knowledge of so many moments is sufficient to reproduce the entire $c$-function, in Fig. \ref{fig:cFuncSGExample} we show the full $c$-function at $\Delta_{\beta}=-1/2$ and $\Delta_\beta=-8$ from LCT with Pad\'e improvement compared to the result from integrability.  Again, the agreement is excellent.

\section{Formulation in terms of UV CFT data}
\label{sec:CFTFormulation}

Up to this point, we have done all computations by relying on the fact that a scalar theory with a $V(\phi)$ potential has a Fock space description for the states and interactions.  Although this is the usual starting point for Lagrangian theories, one of the motivations for using a basis of states constructed in terms of primary operators of the UV CFT is that in principle such a construction can be applied to RG flows that do not have any known Lagrangian description, but instead are defined only in terms of the CFT data of a UV fixed point and its relevant deformations.  Understanding how to treat such deformations in general in a lightcone quantization Hamiltonian formulation is still an open problem, but in the special case of a single relevant deformation $\CO_R$ with $\Delta<1$ in $d=2$ dimensions, a procedure was given in \cite{Fitzpatrick:2018ttk,Chen:2023glf,Fitzpatrick:2023aqm}.  The final result was that the ``effective''  Hamiltonian for lightcone quantization is 
\begin{equation}
H_{\rm eff} = \frac{\lambda_{\rm LC}}{2\pi}  \int dx^- \CO_R(x) \CO_R(\infty), \quad \lambda_{\rm LC} = \lambda_{\rm ET} \< \CO_R\>_{\rm ET},
\label{eq:Heff}
\end{equation}
where $\CO_R(\infty)$ is $\CO_R$ mapped to conformal infinity, which takes on an unambiguous definition when the matrix elements of $H_{\rm eff}$ are evaluated between our basis states created by Fourier transforms primary operators.

  If $V(\phi)$ is of the form $\cos(\beta \phi)$ with $0 < \beta < \sqrt{4\pi}$, or $\beta$ pure imaginary, then $\Delta_\beta <1$ and so the effective lightcone Hamiltonian formulation (\ref{eq:Heff}) should apply here as well.  However, this formulation looks very different from the Fock space formulation of the matrix elements of $V(\phi)$, and it is not at all obvious that they should agree.  In fact, we have checked explicitly that these two methods do exactly agree in this case!  In this section, we will show mechanically how, from very different starting points, they ultimately reduce to the same computation.

\subsection*{Equivalence of algorithms at $\Delta_\beta<1$}\label{sec:equivalence}

When we compute the matrix elements of the lightcone Hamiltonian,  the bra and ket states are manifestly the same in both cases, since they are simply Fourier transforms over primary operators $\CO_i$ constructed from sums and products of derivatives $\partial^k \phi$ of $\phi(x)$.  Their matrix elements are therefore Fourier transforms of correlation functions of these operators together with the Hamiltonian:
\begin{equation}
\< \CO_i, p | P_+ | \CO_j, p'\> = \int d^2 x d^2 x' dy^- e^{i (p \cdot x - p'\cdot x')} \< \CO_i(x) \CV(y^-) \CO_j(x')\>, \quad P_+ = \int dy^- \CV(y^-).
\end{equation}
The difference between the formulations comes entirely from the treatment of the interaction term $\CV$.  In the Fock space formulation, $\CV(y^-)$ is 
\begin{equation}
\CV(y^-) \sim \cos\beta \phi(y^-),
\label{eq:FockV}
\end{equation}
 evaluated by series expanding $\cos \beta \phi$ in powers of $\phi$ and expressing each $\phi$ in terms of creation and annihilation operators.  By contrast, in the CFT formulation, $\CV(y^-)$ is 
 \begin{equation}
 \CV(y^-) \sim \cos(\beta \phi(y^-)) \cos (\beta \phi(\infty)),
 \label{eq:CFTV}
 \end{equation}
  evaluated by computing the CFT four-point function $\< \CO_i(x) \cos(\beta \phi(y_1)) \cos \beta(\phi(y_2)) \CO_j(x')\>$ and conformally mapping $y_2$ to $\infty$.  So the main thing to understand is why the three-point functions
\begin{equation}
\< \CO_i(x) \CV(y^-) \CO_j(x')\>
\end{equation} 
are the same in both cases. 

 At first sight, it would seem impossible for them to be identical, since the former is missing a factor of $\cos(\beta \phi(\infty))$ relative to the second one.  Somehow, the point at infinity must sneak in implicitly in the evaluation of $\cos(\beta \phi)$ in the Fock space formulation.  And in fact, it does, as follows.  Whenever we evaluate  a correlator from the series expansion of $\cos \beta \phi$, we first compute the matrix element
\begin{equation}
\< \CO_i | \phi^n(y) | \CO_j\>
\end{equation}
using the radial quantization representation of the bra and ket states.  Then, to restore the position dependence of $\CO_i(x)$ and $\CO_j(x')$, we have to perform a conformal transformation.  Since $\phi$ is not a primary operator, it transforms in a complicated way under these conformal transformations, essentially as a nonlocal operator.  More precisely, we first evaluate $\prod_{i=1}^n \partial\phi(y_i)$, since $\partial \phi$ is a primary operator, and then we integrate $\partial \phi(y_i)$ to get $\phi(y_i)$.  The point at $\infty$ enters through the evaluation of this integral, which in practice must be $\int_{\infty}^{y_i} dy' \partial \phi(y')$ in order reproduce the fact that the correlator vanishes at large $y_i$. 

Now we will demonstrate how this works at a more detailed level.  In both methods, we define the primary operators as sums over powers of derivatives of the form $\partial^k \phi$:
\begin{equation}
\CO_i(x) = \sum_{\vec{k}} c_{\vec{k}} \partial^{\vec{k}} \phi, \qquad \partial^{\vec{k}} \phi \equiv \partial^{k_1}\phi \dots\partial^{k_n}\phi.
\end{equation}
In radial quantization, the field $\phi$ has the following mode expansion
\begin{equation}
\partial \phi(z) = \frac{i}{\sqrt{4\pi}} \sum_{k=1}^\infty \sqrt{k} (z^{-k-1} a_k + x^{k-1} a_k^\dagger), \qquad [a_k, a^\dagger_{k'}] = \delta_{k,k'}.
\end{equation} 
Because $\partial \phi$ is the spin-1 conserved current, these creation and annihilation operators are also  generators $J_n$ of the Kac-Moody algebra:
\begin{equation}
 \partial \phi(z) \propto J(z) = \sum_{k=-\infty}^\infty \frac{J_k}{z^{k+1}}, \quad J_k = \sqrt{|k|} \left\{ \begin{array}{cc} a_k & k>0 \\ a^\dagger_{-k} & k<0 \end{array} \right\}.
\end{equation}
Each bra states $\< \CO_i|$ can therefore be written as a sum over terms, where each term is a product of $a_k \sim J_{k}$ factors.  In the first representation (\ref{eq:CFTV}), we move $J_k$ factors from the left to the right by commuting them past the two $\cos \beta \phi$ factors. First expand $\cos \beta \phi = \frac{1}{2} (e^{i \beta \phi} + e^{-i \beta \phi})$; then by charge conservation, only the two cross-terms $e^{i \beta \phi(y_1)} e^{-i \beta \phi(y_2)}$ and its Hermitian conjugate contribute to the four-point function. Moving a factor of $J_k$ past these terms involves the following commutator:
\begin{equation}
 [\frac{J_k}{\sqrt{|k|}}, e^{i \beta \phi(y_1)} e^{-i \beta \phi(y_2)}] = e^{i \beta \phi(y_1)} e^{-i \beta \phi(y_2)} i \beta  \frac{(y_1^k - y_2^k)}{\sqrt{|k|}}.
\end{equation}
In the four-point function $\< \CO_i | e^{i \beta \phi(y_1)} e^{-i \beta \phi(y_2)} | \CO_j\>$, we can set $y_2=1$ without loss of generality.  Consequently, each commutator involved introduces a factor of 
\begin{equation}
i \beta \frac{y_1^k-1}{\sqrt{|k|}}
\label{eq:CFTfac}
\end{equation}
and after all the $J_k$ are removed by commuting them to the left or to the right, we finally end up with the product of all these factors, times the two-point function
\begin{equation}
\< e^{i \beta \phi(y_1)} e^{-i \beta \phi(y)}\> = \frac{1}{|1-y_1|^{2\Delta_\beta}}.
\end{equation}
For later convenience, let us define the function $F(y_1)$ to be the product of all the factors of the form $i \beta \frac{y_1^k-1}{\sqrt{|k|}}$ produced by evaluating these commutators:
\begin{equation}
F(y) = \left( i \beta \frac{y^{k_1}-1}{\sqrt{|k_1|}} \right)\dots \left( i \beta \frac{y^{k_n}-1}{\sqrt{|k_n|}}\right).
\label{eq:CFTTotalFac}
\end{equation}

By contrast, in the second representation (\ref{eq:FockV}), we move factors $a_k$ from the left to the right by commuting them past the Taylor series of $\cos(\beta \phi)$:
\begin{equation}
\begin{aligned}
[ a_k, \cos(\beta \phi(y))] &= \sum_{n=0}^\infty \frac{(i\beta)^n}{n!} [a_k, \phi^n(y)] = \sum_{n=0}^\infty \frac{(i\beta)^n}{n!} n \phi^{n-1}(y) [a_k, \phi(y)] \\
  &= \cos (\beta \phi(y)) (i \beta) [a_k, \phi(y)].
\end{aligned}
\end{equation}
However, eventually we have to restore the position dependence of the external operators $\CO_i(x), \CO_j(x')$ by doing conformal transformations.  In order to have a simple transformation rule for each term in the series expansion, we first replace each factor $\phi(y)$ with $\partial \phi(y_i)$ (with a different $y_i$ for each factor $\phi$), then we map $\CO_i(\infty)$ to $\CO_i(x)$ and $\CO_j(0)$ to $\CO_j(x')$.  Then we integrate all of the $y_i$s to turn $\partial \phi(y_i)$ into $\phi(y_i)$, and finally set all the $y_i$s to 1.  The result of this procedure is that each commutator $[a_k, \phi(y)]$ produces a factor\footnote{See \cite{Anand:2020gnn} equation (D.10).}
\begin{equation}
[a_k, \phi(y)] \rightarrow \frac{v^k-1}{\sqrt{|k|}}, \qquad v \equiv \frac{x-y}{x'-y}.
\end{equation}
Consequently, each time we evaluate such a commutator, the $\cos \beta \phi$ factor becomes
\begin{equation}
 \cos (\beta \phi(y)) (i \beta)  \rightarrow  \cos (\beta \phi(y)) (i \beta)  (i \beta) \frac{v^k-1}{\sqrt{|k|}},
 \end{equation}
so that, by inspection, each time we pick up a new factor which is exactly the same as the factor (\ref{eq:CFTfac}) in the first method, but with $v$ instead of $y_1$.  Consequently, the product of all these factors is the same function $F$ as in (\ref{eq:CFTTotalFac}):
\begin{equation}
\< \CO_i(x)V(\phi) \CO_j(x')\> = F(v), \quad  V(\phi) =  -\cos (\beta \phi(y)).
\end{equation}

To complete the argument, we need to show that the function $F(y)$ from the first method gives the same contribution to the final lightcone Hamiltonian matrix elements as does the function $F(v)$ from the second method.  In \cite{Fitzpatrick:2023aqm}, it was found that each term of the form $y^m$ in $F(y)$, for any integer power $m$, gives the following contribution to the Hamiltonian:
\begin{equation}
F = \sum_m a_m y^m \Rightarrow P_+ =2\pi \frac{\sqrt{\Gamma(2\Delta_i)\Gamma(2\Delta_j)}}{\Gamma(\Delta_i + \Delta_j-1)} \sum_m a_m |m|.
\label{eq:CFTrule}
\end{equation}

On the other hand, in \cite{Anand:2020gnn}, it was shown that any $F(v)$ could be decomposed into a sum of terms of the form
\begin{equation}
F(v) = \sum_{k,k'} a_{k,k'} (v^k-1)(1-v^{-k'}),
\end{equation}
with $k$ and $k'$ positive, and that  each such term contributes to the Hamiltonian as follows:
\begin{equation}
P_+ = \pi \frac{\sqrt{\Gamma(2\Delta_i)\Gamma(2\Delta_j)}}{\Gamma(\Delta_i + \Delta_j-1)}  \sum_{k,k'} a_{k,k'} \min(k,k').
\end{equation}
To see the connection between these expressions, decompose $(v^k-1)(1-v^{-k'})$ into a sum over powers of $v$ and apply the rule (\ref{eq:CFTrule}) to each power:
\begin{equation}
\begin{aligned}
&
 (v^k-1)(1-v^{-k'-1}) = 
 (v^k + v^{-k'}-v^{k-k'}-1)\\
&  \stackrel{(\ref{eq:CFTrule})}{\rightarrow} \pi  \frac{\sqrt{\Gamma(2\Delta_i)\Gamma(2\Delta_j)}}{\Gamma(\Delta_i + \Delta_j-1)} 
 (|k|+ |k'| - |k-k'|-0)  
 = \frac{\sqrt{\Gamma(2\Delta_i)\Gamma(2\Delta_j)}}{\Gamma(\Delta_i + \Delta_j-1)} 
  \min(k,k'),
\end{aligned}
\end{equation}
which therefore exactly reproduces the Fock space method prescription.

\section{\texorpdfstring{$\Delta_\beta=1$}{Delta = 1}: Free Fermion Limit }
\label{sec:FreeFermion}

The sine-Gordon model is dual to a fermion with a quartic interaction term. As $\beta$ approaches $\sqrt{4\pi}$, $\Delta_\beta$ approaches $1$.  At this point, the coefficient of the quartic term vanishes and the theory becomes dual to a free fermion.  Previously, in Figure \ref{fig:SineGordonCFunc}, we compared the full $c$-function of sine-Gordon at $\Delta_\beta$ to that of a free fermion massive, with excellent agreement.

In fact, in the case $\Delta_\beta =1$, we will be able go much farther than making this kind of numeric comparison, and will actually see explicit analytic agreement at the level of the lightcone Hamiltonians between the free fermion and the sine-Gordon model.  As mentioned above, what makes $\Delta_\beta=1$ special from the perspective of the lightcone Hamiltonian is that the effective lightcone parameter $\lambda_{\rm LC}$ diverges at $\Delta_\beta=1$, because the expectation value $\< \cos \beta \phi\>$ diverges there.  This divergence is a consequence of the fact that when $\Delta_\beta=1$, there is a logarithmic divergence at second order in perturbation theory in the equal-time coupling $\lambda_{\rm ET}$. By treating $\Delta_\beta$ as a free parameter this divergence is effectively regulated by dimensional regularization and converted into a $1/(\Delta_\beta-1)$ pole:
\begin{equation}
\< \cos \beta \phi\>  = \mu \frac{\pi}{1-\Delta} + \textrm{regular}. \qquad  \mu= \lambda_{\rm ET}/4\pi.
\end{equation}
Consequently, the lightcone coupling $\lambda_{\rm LC}$ also diverges in this limit.  Naively, one would therefore expect that the LCT mass spectrum would diverge at any finite truncation, and at best one could  approach  closer and closer to $\Delta_\beta=1$ with larger truncations.  Remarkably, what actually happens is that even at finite truncation, the lightcone Hamiltonian has exact null vectors when $\Delta_\beta=1$.  For fixed $\lambda_{\rm LC}$, the mass-squareds of these null vectors vanish linearly in $\Delta_\beta-1$,
\begin{equation}
\mu_n^2 \propto \lambda_{\rm LC} (\Delta_\beta-1).
\end{equation}
Upon substituting the relation for $\lambda_{\rm LC}$ in terms of $\lambda_{\rm ET}$ and the vev $\< \cos \beta \phi\>$, this $\Delta_\beta-1$ factor cancels the $1/(\Delta_\beta-1)$ pole in the vev, leading to a finite mass for these states.  By contrast, states with nonzero masses as $\Delta_\beta \rightarrow 1$ with $\lambda_{\rm LC}$ held fixed become infinitely heavy and decouple from the theory.
This pattern of masses lifting at $\Delta_\beta \rightarrow 1$ is reminiscent of the fact that most states in the LCT basis become infinitely heavy for a free fermion in the presence of a mass term $\psi \frac{1}{\partial_-} \psi$, due to the singularity at zero momentum from  $1/\partial_-$ \cite{Anand:2020gnn}.  That is, in the theory of a free massive fermion in an LCT basis, there are special `Dirichlet' states whose wavefunctions vanish when any of the individual fermion momenta goes to zero,\footnote{This is the reason for the name `Dirichlet', since in momentum space the condition is a Dirichlet condition at the kinematic boundaries.} in order to cancel the $1/p_-$ singularity coming from the mass term.  The similarity between these two descriptions  is not a coincidence, and we will explicitly see that the Dirichlet fermion states map, under bosonization, to the states that remain light at $\Delta_\beta\rightarrow 1$ in the bosonic description.

We begin with the simplest massless mode, which is made from a linear combination of the operators $\partial \phi$ and $(\partial \phi)^3$.  We find that the LCT Hamiltonian for these two bosonic states is
\begin{equation}
P^2 =2  \lambda_{\rm LC}  \left( \begin{array}{cc}  1 & \sqrt{5} \\ \sqrt{5} & 5 \end{array} \right),
\end{equation}
which has a null vector
\begin{equation}
\propto | \partial \phi, p\> - \frac{1}{\sqrt{5}} | (\partial \phi)^3 , p \>
\label{eq:DirMassless}
\end{equation}
The relative sign here is convention-dependent since it can be removed by redefining $ | (\partial \phi)^3 , p \> \rightarrow -  | (\partial \phi)^3 , p \>$.  

Now we will see  how this state is reproduced from the Dirichlet basis in the free fermion description. The ``Dirichlet'' basis is picked out by the condition that all factors of the free fermion operator $\psi$ must have at least one derivative acting on them, in order to compensate for the $\frac{1}{\partial}$ divergence from the mass term.  The lowest-dimensional neutral operator satisfying this condition is  $\partial \psi^* \partial \psi$, which we need to map under bosonization into operators in the bosonic description.  We will use a convention where $\phi$ and $\psi$ are normalized as $\< \phi(x) \phi(0)\> = \log x$ and $\< \psi^*(x) \psi(0) \> = x^{-1}$.  

To begin, note that (in both the free fermion or the free scalar description, since they are duals of each other) there is exactly one primary operator at dimension 1 and  one at dimension 3, which we can  call  $\CO_1$ and $\CO_3$. We adopt the following normalization for both $\CO_1$ and $\CO_2$:
\begin{equation}
\< \CO(x) \CO(0)\> = x^{-2 \Delta_\CO}
\end{equation}
With this normalization, one can uniquely (up to a sign) fix $\CO_1$ and $\CO_2$ in both descriptions by the condition that they are primary operators with dimensions 1 and 3, respectively:
\begin{equation}
\CO_1 = \partial \phi = \psi^* \psi, \qquad \CO_3 = \frac{(\partial \phi)^3}{\sqrt{6}} = \frac{4 \partial \psi^* \partial \psi  - \partial^2 \psi^* \psi -  \psi^* \partial^2 \psi}{\sqrt{24}}.
\end{equation}
By inspection,
\begin{equation}
6 \partial \psi^* \partial \psi = \partial^2 \CO_1 + \sqrt{24} \CO_3.
\label{eq:dpsidpsiToBoson}
\end{equation}
Any operator normalized by $\< \CO(x) \CO(0)\> = x^{-2\Delta_\CO}$ creates a normalized momentum-space state $|\CO, p\>$ according to \cite{Anand:2020gnn}
\begin{equation}
| \CO, p \> = \sqrt{\frac{\Gamma(2\Delta_\CO)}{\pi}} \int dx e^{i p \cdot x} \CO(x) | 0\>,
\end{equation}
where we take $p=1$. A short calculation shows that
\begin{equation}
\< (\partial \psi^* \partial \psi)(x)(\partial \psi^* \partial \psi)(0)\> = \frac{4}{(x-y)^6},
\end{equation}
so
\begin{equation}
\begin{aligned}
| \partial \psi^* \partial \psi, p \> &= \sqrt{\frac{\Gamma(6)}{\pi}} \int dx e^{i p \cdot x} \frac{\partial \psi^* \partial \psi(x)}{2} |0\>, \\
| \partial \phi , p \> &= -  \sqrt{\frac{\Gamma(2)}{\pi} }\int dx e^{i p \cdot x} \partial^3 \phi(x) |0\>, \\
| (\partial \phi)^3 , p \> &= \sqrt{\frac{\Gamma(6)}{\pi}} \int dx e^{i p \cdot x}  (\partial \phi)^3(x) |0\>.
\end{aligned}
\end{equation}
Fourier transforming both sides of (\ref{eq:dpsidpsiToBoson}),
\begin{equation}
(2)(6) \sqrt{\frac{\pi}{\Gamma(6)}} | \partial \psi^* \partial \psi, p \> = - \left( \sqrt{\frac{\pi}{\Gamma(2)}} | \partial \phi, p\> - \sqrt{24} \sqrt{ \frac{\pi}{\Gamma(6)}} | (\partial \phi)^3, p\> \right)
\end{equation}
which simplifies to
\be
\sqrt{ \frac{6}{5}} | \partial \psi^* \partial \psi, p \> = - |\partial \phi, p\> + \frac{1}{\sqrt{5}} | (\partial \phi)^3, p\> .
\ee
This exactly reproduces the state we found in (\ref{eq:DirMassless}) in the sine-Gordon Hamiltonian!

 As the dimension of the operators increases, the Dirichlet states can contain more fermions and become more complicated.  However, the lowest-dimensional four-fermion Dirichlet state is $(\partial^2 \psi \partial \psi )(\partial^2 \psi \partial \psi )^*$, which has dimension 8, so up to $\Delta \le 7$ we can just consider the two-fermion Dirichlet states. 
In momentum space, their wavefunctions have simple expressions in terms of normalized Jacobi polynomials:\footnote{See \cite{Anand:2020gnn} for a derivation of these wavefunctions in the Dirichlet basis.}
\begin{eqnarray}
\hat{P}_n^{(\alpha, \beta)}(y) &\equiv& \mu_n(\alpha, \beta) P_n^{(\alpha, \beta)}(y), \nn\\
\mu_n(\alpha, \beta) &=& \sqrt{ \frac{\Gamma(n+1)\Gamma(n+\alpha+\beta+1)\Gamma(2n+\alpha+\beta+2)}{\Gamma(n+\alpha+1)(\Gamma(n+\beta+1)\Gamma(2n+\alpha+\beta+1)}} .
\end{eqnarray}
The Dirichlet two-fermion states are Fourier transforms of local operators like the example $\partial \psi^* \partial \psi$ we saw above.  Their explicit form is
\begin{equation}
D_n(z_2) = \left[ \hat{P}_n^{(2,2)}(\frac{\partial_1 - \partial_2}{\partial_1 + \partial_2}) \partial_1 \partial_2 (\partial_1+\partial_2)^n \psi^*(z_1) \psi(z_2)\right]_{z_1\rightarrow z_2} .
\label{eq:JacobiDirichlet}
\end{equation}
For example, the first two operators are $D_0 \propto \partial \psi^* \partial \psi$ and $D_1 \propto \partial^2 \psi^* \partial \psi - \partial \psi^* \partial^2 \psi$.

An efficient way to translate these operators into the bosonic description is as follows.   First, recall that each fermion in the bosonic description is a vertex operator:
 \begin{equation}
 \psi = : e^{i \phi} : , \qquad \psi^* = :e^{-i \phi}: .
 \end{equation}
We have added explicit normal-ordering symbols $: \dots :$ since their effect will be important momentarily.  To construct a composite operator like $\psi^* \psi$, we can use the following standard contour-integral procedure:
 \begin{equation}
(\psi^*\psi)(z) = \oint_z \frac{dw}{2\pi i } \frac{1}{z-w} \psi^*(w) \psi(z).
 \end{equation}
To apply this formula to the vertex operators, we can use the following general expression for normal-ordered products of vertex operators:
\begin{equation}
: e^{i \phi(w)}: :  e^{-i \phi(z)}:  = \frac{1}{w-z} :e^{i (\phi(w)- \phi(z))} :
\end{equation}
Consider for simplicity a two-fermion Dirichlet state.  In general, it can be written as a sum over terms of the form
\begin{equation}
D(z) = \oint_z \frac{dw}{2\pi i} \frac{1}{z-w} \partial^n \psi^*(w) \partial^m \psi(z).
\end{equation}
This operator takes the following form  in the bosonic description:
\begin{equation}
D(z) = \oint_z \frac{dw}{2\pi i} \frac{1}{z-w} \partial_w^n \partial_z^m \left( \frac{1}{w-z} : e^{i (\phi(w)- \phi(z))}: \right) .
\end{equation}
Evaluating the $\oint \frac{dw}{2\pi i (z-w)}$ integral amounts to series expanding $w$ around $z$ and keeping the constant term, so the exponential can be expanded as a sum over derivatives of $\phi(z)$, and only a finite number of such terms contribute.  For instance, in this way one can efficiently substitute into (\ref{eq:JacobiDirichlet}) and find that $D_0 \propto \partial^3 \phi + 2 (\partial \phi)^3$, and $D_1 \propto (\partial \phi)^4 + (\partial^2 \phi)^2$, $D_2 \propto \frac{28}{5} (\partial \phi)^5 + 16 \partial \phi (\partial^2 \phi)^2 - 6 (\partial^3 \phi) (\partial \phi)^2 - \frac{1}{15} \partial^5 \phi$, etc. We have checked explicitly up to $\Delta \le 7$ that all the null vectors of the lightcone Hamiltonian for the $\cos(\beta \phi)$ potential at $\beta = \sqrt{4\pi}$ exactly match these operators.

Finally, we can go one step farther and check that the Hamiltonians themselves match between the fermionic and bosonic description! In the fermionic description, the mass term for the fermion with mass $M$ in a lightcone Hamiltonian is $P^2 \supset M^2 \psi \frac{1}{\partial} \psi$, and in the Dirichlet basis its matrix elements are given by the following integrals:
\begin{equation}
M^2 \int_0^1 dx x(1-x) \hat{P}_n^{(2,2)}(1-2x) \hat{P}_m^{(2,2)}(1-2x)
\end{equation}
The first few of these are, for $n,m=0,\dots, 4$:
\begin{equation}
M^2 \left(
\begin{array}{ccccc}
 5 & 0 & \sqrt{3} & 0 & \sqrt{\frac{13}{14}} \\
 0 & 7 & 0 & \sqrt{11} & 0 \\
 \sqrt{3} & 0 & 9 & 0 & 3 \sqrt{\frac{39}{14}} \\
 0 & \sqrt{11} & 0 & 11 & 0 \\
 \sqrt{\frac{13}{14}} & 0 & 3 \sqrt{\frac{39}{14}} & 0 & 13 \\
\end{array}
\right)
\label{eq:DirichletMassTermForSG}
\end{equation}
The smallest eigenvalue of this matrix at this truncation is $4.18325 M^2$, and approaches $4 M^2$ as more basis states are included, in agreement with the invariant mass-squared threshold for two-particle states.  
The exact matrix elements are reproduced from the sine-Gordon description as follows.  Since we already have the `Dirichlet' states in the bosonic basis, we can simply evaluate the lightcone Hamiltonian in this basis.  The only subtlety is that the lightcone Hamiltonian exactly at $\Delta_\beta$ is of the form $\frac{\Delta_\beta-1}{\Delta_\beta-1}$, since the $(\Delta_\beta-1)$ in the denominator comes from the vev $\< \cos \beta \phi\>$ and the $\Delta_\beta-1$ in the numerator comes from the fact that the Dirichlet states are massless at $\Delta_\beta=1$ when we hold $\lambda_{\rm LC}$ fixed.  Fortunately, this problem is  resolved by evaluating the Hamiltonian matrix elements on the Dirichlet basis states for $\Delta_\beta <1$ and then taking the limit. For instance, consider the first Dirichlet state (\ref{eq:DirMassless}).  For any value of $\beta$, we can calculate the $P^2$ matrix element for the  $\cos \beta \phi$ potential in lightcone quantization, with the result
\begin{equation}
\< D_0 | P^2 | D_0\> = \frac{\lambda_{\rm LC}}{4\pi} \frac{20 \pi }{3} \Delta_\beta (1-\Delta_\beta)(4-\Delta_\beta).
\end{equation}
Substituting $\lambda_{\rm LC} = \lambda_{\rm ET} \< \cos \beta \phi\>_{\rm ET}$ with (\ref{eq:VevReln}), 
\begin{equation}
\< D_0 | P^2 | D_0\> = M_s^2 \frac{5 \pi  (4-\Delta_\beta ) (1-\Delta_\beta ) \Delta_\beta  \tan \left(\frac{\pi  \Delta_\beta }{4-2 \Delta_\beta }\right)}{3 (2-\Delta_\beta )}  \stackrel{\Delta_\beta \rightarrow 1}{\rightarrow} 5 M_s^2,
\end{equation}
in exact agreement with the first element of $P^2$ in (\ref{eq:DirichletMassTermForSG}).  We have checked up to $\Delta \le 7$ that this exactly reproduces the matrix elements show in (\ref{eq:DirichletMassTermForSG}).

\section{Puzzles at \texorpdfstring{$\Delta_\beta>1$}{Delta > 1}}
\label{sec:Puzzles}

There are a few caveats that must be made for the application of lightcone quantization to $V(\phi)$ theories.  The first such caveat is that the effective parameters in the Lagrangian in lightcone quantization are renormalized compared to the corresponding equal-time parameters, which can be understood as the effect of integrating out lightcone zero modes \cite{Fitzpatrick:2018ttk,Chen:2023glf,Fitzpatrick:2023aqm}.  This caveat has already played a significant role in our analysis of the sine-Gordon and sinh-Gordon models, where the mapping between equal-time and lightcone parameters is known analytically.  We emphasize however that such relations are not strictly necessary for applying lightcone quantization, since one may always choose to define the theory by its lightcone parameters; the mapping is only necessary if one wants to compare a specific point in parameter space in lightcone quantization to a specific point in parameter space in equal-time quantization.  

\begin{figure}[ht]
\centering
\includegraphics[width=0.49\linewidth]{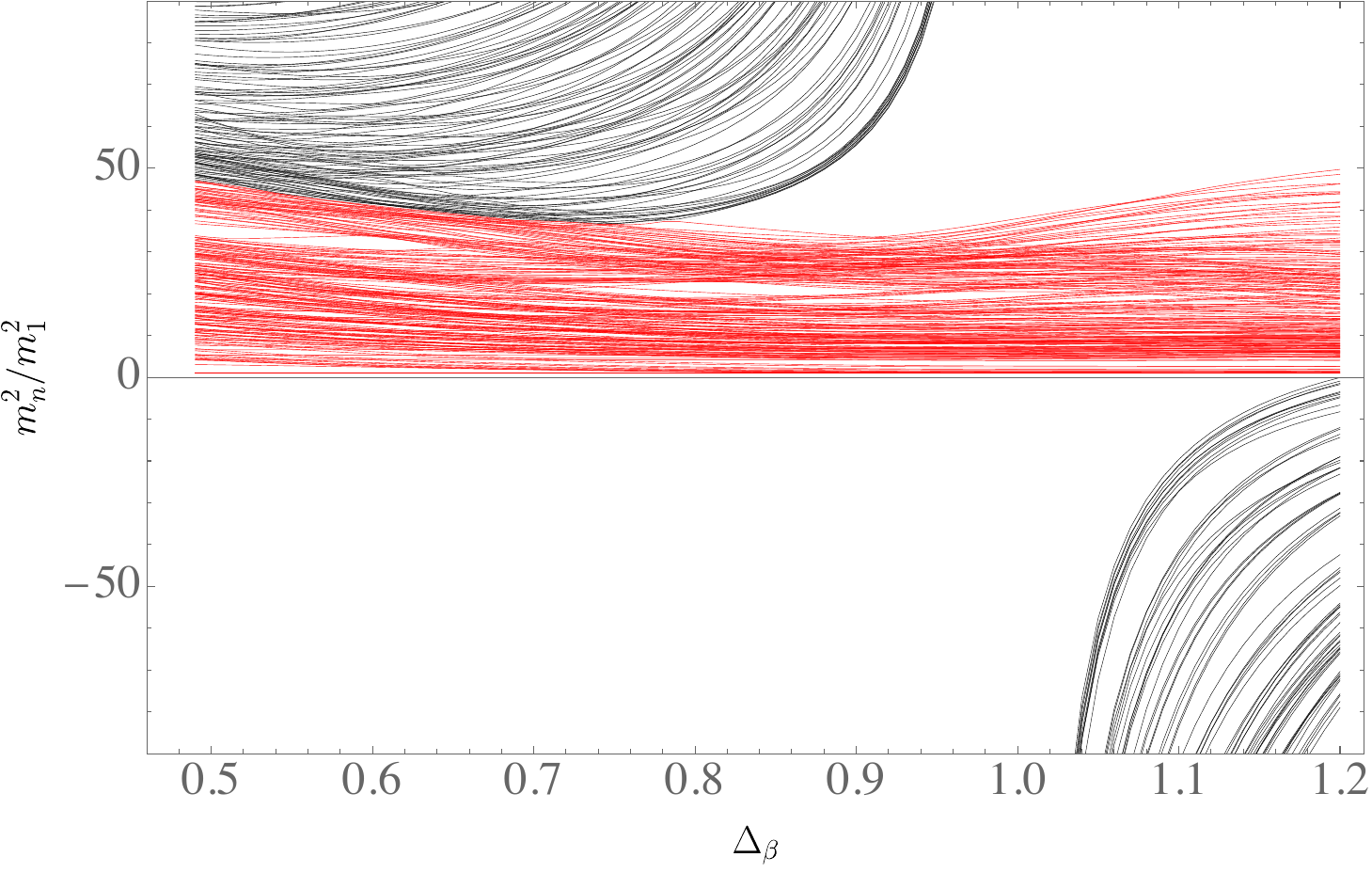}
\includegraphics[width=0.49\linewidth]{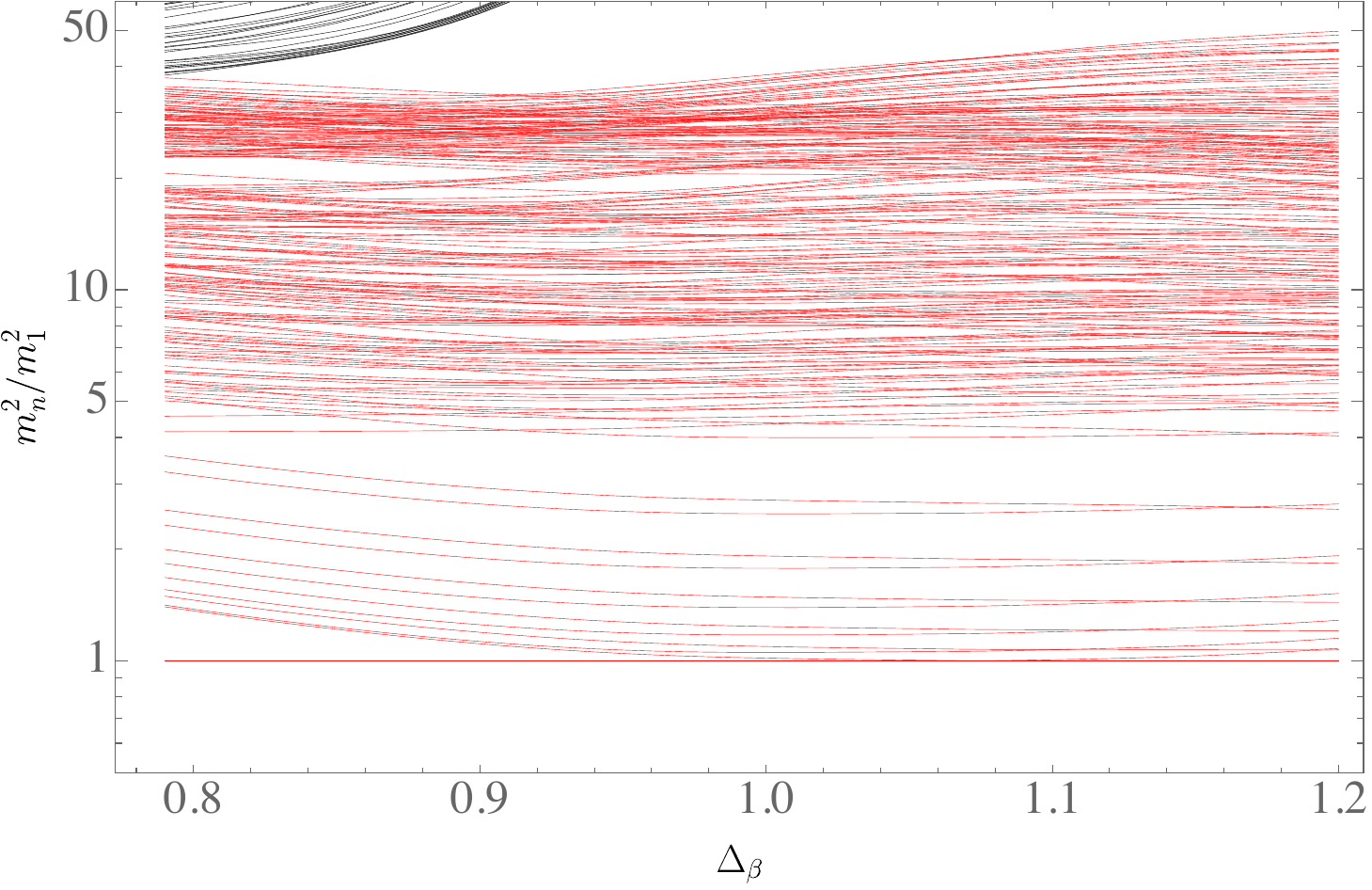}
\caption{Spectrum of the sine-Gordon model passing from $\Delta_\beta<1$ to $\Delta_\beta>1$, at $\Delta_{\rm max}=20$.  As $\Delta_\beta$ approaches 1 from below, a subset of the spectrum (the ``Dirichlet'' subspace, shown in red) maintains a finite mass, while the rest of the states ({\it black}) become infinitely heavy and decouple.  At $\Delta_\beta = 1 +\epsilon$, the heavy decoupling states reenter the spectrum from below, while the Dirichlet states pass through the transition smoothly.  The right plot is a zoomed-in version of the left plot, to show the Dirichlet states in more detail.}
\label{fig:Decoupling}
\end{figure}

A more serious caveat arises when we consider regions of equal-time parameter space which cannot be mapped to any $V(\phi)$ in lightcone quantization.  This situation appears to arise when we consider the sine-Gordon model with $1 < \Delta_\beta < 2$ (equivalently, $\sqrt{4\pi} < \beta < \sqrt{8\pi}$). In Fig.~\ref{fig:Decoupling}, we show the spectrum at truncation $\Delta_{\rm max}=20$ as we pass from $\Delta_\beta<1$ to $\Delta_\beta>1$.  The key salient feature is that a subset of states (the ``Dirichlet'' states as $\Delta_\beta \rightarrow 1$, discussed in section \ref{sec:FreeFermion}) pass through the value $\Delta_\beta=1$ smoothly, while the remaining states first approach infinite energy and decouple as $\Delta_\beta \rightarrow 1$ from below, and then reenter the spectrum at negative values for $\Delta_\beta >1$.  

A natural guess is that some states should continue to be lifted out of the spectrum not just at $\Delta_\beta=1$ but also at $\Delta_\beta>1$.  One argument in favor of this possibility comes from the derivation from \cite{Fitzpatrick:2023aqm} of the lightcone effective action (\ref{eq:Heff}), which starts with equal-time quantization and integrates out modes that become heavy in the limit of taking the large momentum limit.  The advantage of this method is that it gives a procedure that is in principle well-defined for obtaining the effective lightcone Hamiltonian, though in practice it may be computationally challenging.  Nonetheless, one can numerically look at the resulting effective lightcone Hamiltonian for large values of the momentum, and it contains contributions that scale like $p^{\Delta_\beta-1}$.  Consequently, these extra pieces can be discarded for $\Delta_\beta<1$, but for $\Delta_\beta>1$ they diverge in the large momentum limit, and therefore will give infinite energies to some subspace of the full Hilbert space.  It seems likely that this is also related to the fact that at $\Delta_\beta>1$, the renormalization of the vacuum energy diverges with the cutoff like $\Lambda^{2(\Delta_\beta-1)}$ in perturbation theory.   Understanding the proper treatment of deformations with $\Delta>1$ is an important open problem in lightcone quantization, but outside the scope of this paper.

\section{Future Directions}
\label{sec:Future}
With the addition of the computational techniques introduced in this paper, it is now possible to do efficient, high accuracy computations of a large class of 2d scalar theories with a $V(\phi)$ potential at strong coupling.  To demonstrate this, we have applied this method to the sine-Gordon and sinh-Gordon models in 2d, improving on previous results as well as probing regions that have proven difficult to study with other methods.  In particular, we obtained accurate results on both sides of the self-duality of the sinh-Gordon model, and provided compelling evidence in favor of the duality.  The reason we were able to probe both sides of the duality is due to significant simplifications that occur in lightcone quantization, which eliminates challenges associated with the zero mode of the scalar field.  These challenges are not entirely invisible in lightcone quantization, but instead get absorbed into the behavior of effective couplings in the lightcone Hamiltonian, which is properly thought of as an effective theory. Moreover, in this case there is a simple conjecture for the exact behavior of the effective lightcone coupling, based on arguments in \cite{Burkardt:1992sz,Chen:2023glf,Fitzpatrick:2023aqm}. The results in this paper numerically confirm this conjecture in the sinh-Gordon and sine-Gordon models at high precision.

It is our hope that the availability of this tool encourages future studies of strongly coupled models using Hamiltonian truncation in lightcone quantization.  The class of 2d scalar models with a $V(\phi)$ is quite large, and there are many interesting cases to pursue.  The most well-known among these include the Landau-Ginzburg description of the RG flow to the minimal model CFTs \cite{zamolodchikov1986conformal,Kastor:1988ef,Coser:2014lla,Lencses:2024wib,Katsevich:2024jgq}. However, we wish to add some cautionary words about the subtleties of studying theories with spontaneous symmetry breaking in lightcone quantization.  Take, for instance, the case of the RG flow to the 2d Ising model, using a quadratic plus quartic potential $V(\phi) = \frac{1}{2}m^2 \phi^2 + \frac{\lambda}{4!} \phi^4$ tuned to the critical point.  It is well-known that the critical value of the dimensionless coupling $\lambda/m^2$ is different in lightcone quantization versus in equal-time quantization, due to the effect of integrating out zero modes.  This issue is fairly trivial to resolve, since it just involves a change in the couplings of the model.  Things become more subtle if one wants to study the low-temperature phase of the model, at couplings where the order parameter $\phi$ gets a nonzero vev $v$.  In this case, it is no longer sufficient to simply modify the coupling $\lambda$.  Rather, in lightcone quantization, when the vacuum undergoes this kind of change it is necessary to expand the lightcone Lagrangian around the new vev of $\phi$ from the very beginning, $\phi \rightarrow v+\phi$, which generates a cubic term $\sim \phi^3$ in the effective action.  Finding the correct coefficient of this cubic term, relative to the quadratic and quartic terms, is a strongly coupled problem that must be solved by analyzing the spectrum in detail.  Such an analysis that determines the cubic coupling should be feasible, but it makes the models less straightforward than the high-temperature phases and one should be aware of the challenges.  Similar considerations apply to  other models  with spontaneous symmetry breaking.

\paragraph{Invitation: Schwinger Model}

\begin{figure}[ht!]
\centering
\includegraphics[width=0.55\linewidth]{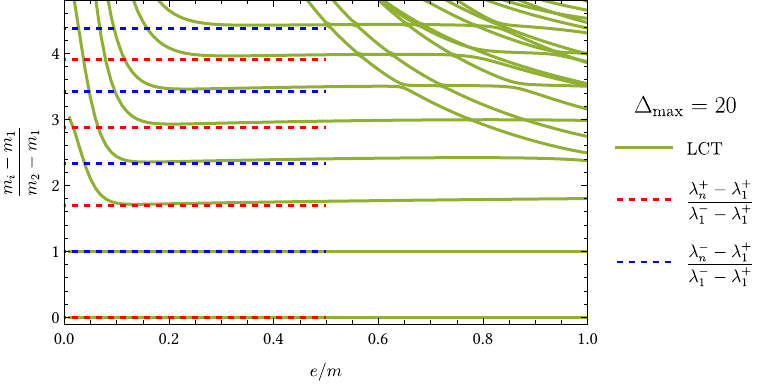}
\includegraphics[width=0.415\linewidth]{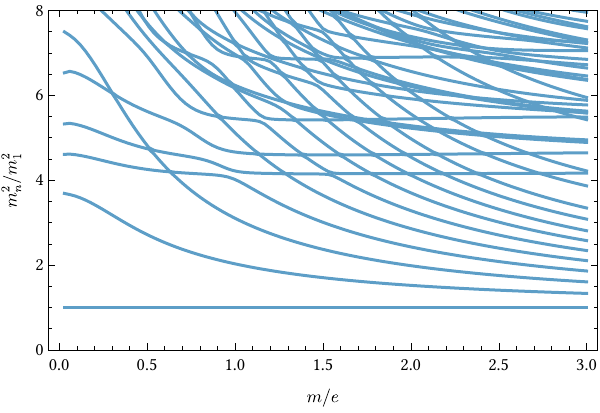}
\caption{\label{eq:fig:Schwinger}
{\bf Left panel:} 
The spectrum of the Schwinger model as a function of the dimensionless parameter $e/m$. The difference between the first and second bound state is normalized to 1, and the lowest bound state energy is subtracted. 
The green solid lines represent The LCT result, and the red (blue) dashed line represent the theorectical prediction of $e\rightarrow 0$ spectrum of symmetric (anti-symmetric) 2-fermion bound states.
In the $e\rightarrow 0$ limit, the theoretical prediction is given (\ref{schwinger-spectrum}). In the plot's subtraction and normalization scheme, the prediction of the spectral lines are $\frac{\lambda_n^{\pm} - \lambda_1^{+}}{\lambda_1^{-}-\lambda_1^{+}}$. We see for small $e/m$, the LCT spectrum first approach the prediction, and diverge for extremely small $e/m$ due to truncation effect.
{\bf Right panel:} The spectrum of Schwinger model for large $e/m$. Most of the bound states disappear into the multi-particle threshold.
}
\end{figure}

One model in particular that is interesting and well-suited to these techniques is the massive Schwinger model \cite{PhysRev.128.2425,Coleman:1975pw} with zero theta angle $\theta$.  It has a single dimensionless ratio of parameters, the charge to mass ratio $e/m$, and is solvable in the limit that either $e$ or $m$ vanishes, but not in between.    In Fig.~\ref{eq:fig:Schwinger}, we show the spectrum of the theory, obtained with our method applied to the model for a range of parameters $e/m$. %at small $e/m$, 
At small $e/m$, the spectrum precisely matches that of bound states formed by non-relativistic particles in a linear potential, where the binding energy is given by the zeros of Airy function and its derivative \cite{hamer1977lattice}
\begin{equation}\label{schwinger-spectrum}
\begin{aligned}
E_{n}^{\pm} &= 2m - \lambda_n^{\pm} \left(\frac{e^4}{4m} \right)\\
{\rm Ai}'(\lambda^{+}_n) &= 0, \quad \lambda^{+}_n = -1.01879,~-3.24820\cdots \\ 
{\rm Ai}(\lambda^{-}_n) &= 0, \quad \lambda^{-}_n = -2.33811,~-4.08795\cdots~.
\end{aligned}
\end{equation}
The $\pm$ means the 2-particle wave function is symmetric (anti-symmetric).
At large $e/m$ the bound states disappear into the multi-particle threshold, qualitatively agreeing with the expectation that the massless Schwinger model is dual to free scalar theory with potential $\phi^2$ \cite{coleman1975quantum}, though in this limit there is no theoretical bound state spectrum to compare to. We believe the quantitative match between the $V(\phi)$ potential and the Schwinger model should hold in all parameter space, not only for the spectrum but also other observables. It would be interesting to  compute more observables of this model and compare with other numerical techniques such as other Hamiltonain truncation methods \cite{Schmoll:2023eez,Ingoldby:2024fcy}, Density Matrix Renormalization Group  \cite{PhysRevD.66.013002,Dempsey:2022nys,Angelides:2023bme,ArguelloCruz:2024xzi,Fujii:2024reh} or lattice Monte Carlo \cite{creutz1995quark,hoferichter1998dynamical,Durr:2003xs,Ohata:2023gru,Ranft:1982bi,Schiller:1983sj,Ohata:2023sqc}.

\begin{center} {\bf Acknowledgments} \end{center}

\noindent We are grateful to Hongbin Chen 
for collaboration in the early stages of this project, and to Erick Arguello, Grigory Tarnopolsky, and  Matthew Walters for helpful discussions.
%, and to ... for comments on a draft. 
ALF and EK are supported by the US Department of Energy
Office of Science under Award Number DE-SC0015845. YX is supported by the Simons Foundation Grant No. 994316 (Simons Collaboration on Confinement and QCD Strings).

\newpage 
\bibliographystyle{JHEP}
\bibliography{refs}

\end{document}